\documentclass[11pt]{article}
\usepackage{arxiv}
\usepackage{geometry}                
\usepackage{graphicx,bm}
\usepackage{epstopdf, epsfig}
\usepackage{xcolor,color}
\usepackage{amsmath}
\usepackage{mathtools}
\usepackage{amssymb}
\usepackage{amsthm}

\newcounter{defcounter}
\newenvironment{myequation}{%
\addtocounter{equation}{-1}
\refstepcounter{defcounter}

\begin{equation}}{\end{equation}}
\usepackage{hyperref}
\usepackage{cleveref}
\usepackage{comment}
\usepackage[color=green!20]{todonotes}

\newcommand{\Blue}[1]{{\color{blue}#1}}

\usepackage[backend=bibtex,firstinits=true,style=numeric-comp,date=year,sorting=none]{biblatex}
\renewbibmacro{in:}{}
\AtEveryBibitem{\clearlist{language}}
\title{Curvature-driven feedback on aggregation-diffusion of proteins in lipid bilayers}
\author{Arijit Mahapatra,
David Saintillan$^*$,
and Padmini Rangamani$^*$ \\
Department of Mechanical and Aerospace Engineering, University of California San Diego,\\
9500 Gilman Drive, La Jolla, CA 92093, U.S.A.\\
Email addresses of corresponding authors: 
\href{dsaintillan@eng.ucsd.edu}{\color{blue}{dsaintillan@eng.ucsd.edu}};
\href{prangamani@ucsd.edu}{\color{blue}{prangamani@ucsd.edu}}
}

\begin{document}

\maketitle

\begin{abstract}
Membrane bending is an extensively studied problem from both modeling and experimental perspectives because of the wide implications of curvature generation in cell biology.
Many of the curvature generating aspects in membranes can be attributed to interactions between proteins and membranes. 
These interactions include protein diffusion and formation of aggregates due to protein-protein interactions in the plane of the membrane. 
Recently, we developed a model that couples the in-plane flow of lipids and diffusion of proteins with the out-of-plane bending of the membrane.
Building on this work, here, we focus on the role of explicit aggregation of proteins on the surface of the membrane in the presence of membrane bending and diffusion. 
We develop a comprehensive framework that includes lipid flow, membrane bending, the entropy of protein distribution, {along with} an explicit aggregation potential and derive the governing equations {for the coupled system}.
We compare this framework to the Cahn-Hillard formalism to predict the regimes in which the proteins form patterns on the membrane. 
We demonstrate the utility of this model using numerical simulations to predict how aggregation and diffusion,  {when} coupled with curvature generation, can alter the landscape of membrane-protein interactions. 
\end{abstract}

\section{Introduction}
\begin{figure}[htbp]
    \centering
    \includegraphics[width=\textwidth]{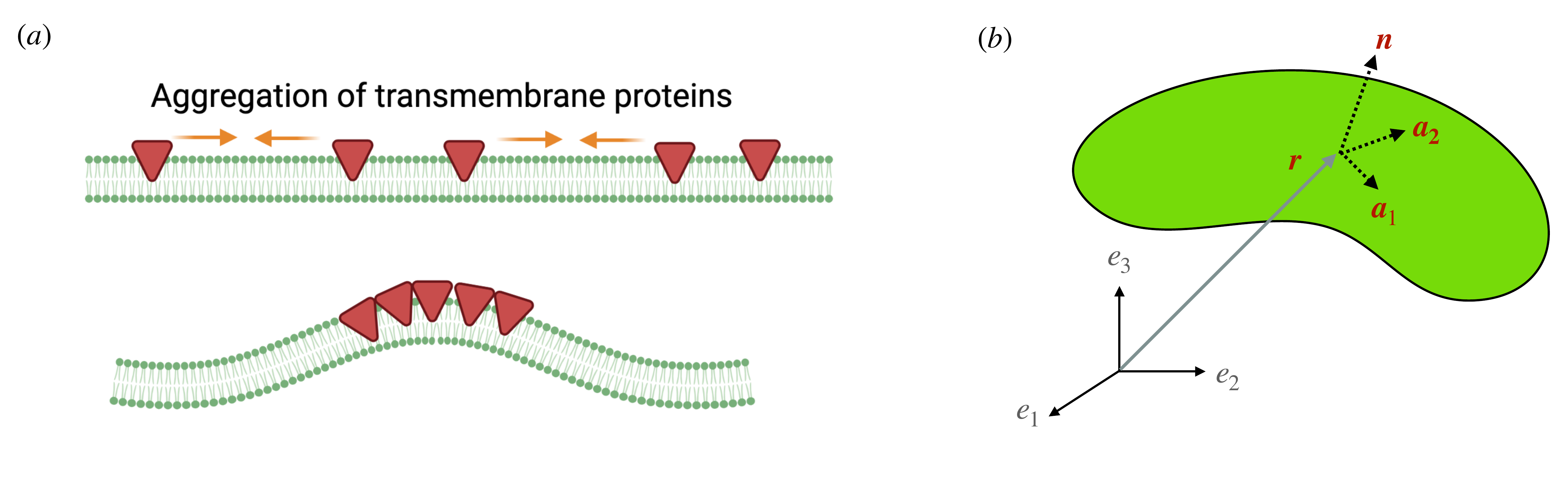}
    \caption{Schematic of protein aggregation and representation of a membrane surface. (\textit{a}) Aggregation of  transmembrane proteins on the membrane can lead to domain formation and curvature generation. Here, we develop a continuum model that captures these different interactions. (\textit{b}) Representation of a membrane surface and the surface coordinates. $\boldsymbol{r}$ is the position vector, $\boldsymbol{a}_1$ and $\boldsymbol{a}_2$ are the tangent basis vectors, $\boldsymbol{n}$ is the unit surface normal.} \label{fig:schematic}
\end{figure}

Cellular membranes contain a variety of integral and peripheral proteins whose spatial organization has biophysical implications for cellular function \cite{cremades2018appl_disease,mukherjee2017gen_appln}. 
In the plane of the membrane, many of these proteins are known to diffuse \cite{kahya2004diff}, induce curvature in the bilayer \cite{mcmahon2005curv_gen}, 
and aggregate either through protein-specific interactions \cite{douglas2010interplay} 
or due to membrane curvature \cite{reynwar2007aggregation}.
Interactions between proteins can also lead to the formation of protein microdomains depending on the strength of interaction forces \cite{weber2019elife1,reynwar2007aggregation}. 
The ability of these proteins to induce curvature, coupled with the ability of curvature to influence the lateral diffusion-aggregation dynamics, can result in a feedback loop between membrane curvature and protein density on the surface (\Cref{fig:schematic}a) \cite{nitschke2012finite,klaus2016curLB,gera2017cahn}.
In addition to protein aggregation, in-plane viscous flow of the lipid molecules has been found to dominate some of the phase-transition kinetics of vesicle shapes \cite{noguchi2004fluid}.
Recently, we showed that the interaction of membrane bending, protein diffusion, and lipid flow can lead to an aggregation-like configuration on the membrane {under specific conditions} \cite{mahapatra2020transport}.


The aggregation of particles in solvents is a well-studied theoretical problem.
Flory \cite{flory1942thermodynamics} and Huggins \cite{huggins1942some} presented a theoretical formulation for a polymer chain in solution and established the conditions that can lead to its phase separation from the solvent. 
In binary alloy systems, there has been significant progress on the modeling of the phase transition mechanisms starting from the fundamental Ginzburg-Landau energy \cite{cahn1961spinodal} that models the interaction energy between the phases as an algebraic expansion in the area fraction of the binary phases around a reference value.
Additionally, there are a number of studies that considered the effect of surface tension in the phase separation of solid solutions with an elastic field as a function of concentration field of solute
\cite{onuki1989long,onuki2001Bin,cahn1962spinodal}.

While the classical theories were developed for three-dimensional continua, domain formation and phase separation on two-dimensional surfaces such as lipid bilayers have been of paramount interest recently. 
The aggregation of proteins on the membrane surface can be viewed as an example of a binary system with lipids and proteins as two phases in a two-dimensional curvilinear space.
For example, a recent modeling study showed that in a reaction-diffusion system, a pair of activator and inhibitor molecules can lead to an aggregation instability in a specific parameter space, and this instability governs the pattern formation of proteins on membranes \cite{Stolerman2020-yl}.
There are many models in the literature that investigate various aspects of phase separation on surfaces. Gera and Salac \cite{gera2017cahn} numerically solved a Cahn-Hilliard system for aggregation-diffusion on a closed torus and observed the temporal evolution of the formation of the aggregation patches. 
In this case, the surface geometry was fixed. 
In a subsequent study, they analyzed the effect of bulk shear flow on the dynamics of the density distribution of species on a deformable vesicle, where the material properties are dependent on the species concentration \cite{gera2018three}. 
Nitschke \textit{et al.} \cite{nitschke2012finite} modeled aggregation-diffusion of a two-phase mixture on a spherical surface with in-plane viscous flow, and presented numerical results on pattern formation between the two phases and its strong interplay with the surface flow.
The relative interactions between the proteins on the cellular membrane can lead to phase segregation and form protein domains depending on the strength of interaction forces compared to the entropy of mixing \cite{weber2019elife1}. 
Such aggregation phenomena have been modeled as a polymerization reaction with a very weak free energy of polymerization \cite{weber2019elife1}. 

Coupling these aggregation phenomena on the membrane surface with membrane deformation is a difficult mathematical and computational problem.
Reynwar \textit{et al.}~\cite{reynwar2007aggregation} modeled the interaction between proteins with the help of an inter-particle energy and showed that curvature alone can lead to aggregation of these protein particles.  
A majority of the aggregation studies in the continuum realm consider an aggregation-diffusion chemical potential, which results in the well-known Cahn-Hilliard equation that represents phase separation.
The energy potential used in studies of protein aggregation on membrane surfaces consists of an inter-molecular aggregation energy and a diffusion potential comprising of the entropy of the protein distribution.
Veksler and Gov \cite{veksler2007phase} considered the Ginzburg-Landau energy potential for the aggregation-diffusion energy and modeled the curvature-diffusion instability to identify the parameter space where such instability occurs.
Mikucki and Zhou \cite{mikucki2017curvature} {developed} a numerical solution for aggregation-diffusion of proteins with bending of the membrane and inviscid flow of lipids.
However, their model assumes that the local membrane curvature is a function of the density of the proteins as opposed to using a spontaneous curvature, resulting in a weak coupling between bending and diffusion. 
Givli \textit{et al.} \cite{givli2012stability} presented a theoretical model of diffusion-aggregation in a multicomponent inviscid stretchable membrane coupled with the bending of the membrane. 
Additionally, they performed a stability analysis of the system on a sphere, and obtained the most critical modes for the instabilities. 



While the models described above capture different aspects of the same problem, here, we sought to develop a comprehensive mathematical model that captures the coupled diffusion and aggregation dynamics, where the proteins induce a curvature resulting in membrane bending and lipids can flow in the plane of the membrane.
Such a framework can allow us to explore how the different transport contributions in the plane of the membrane (protein aggregation, protein diffusion, and lipid flow) can contribute both to the formation of protein microdomains and to the curvature generation capability of the membrane.
The manuscript is organized as follows.
The full system of governing equations is presented in Section~\ref{sec:model}.  
We first analyzed the special case in the absence of bending and reduced the model to a classic Cahn-Hilliard system in Section~\ref{sec:CH}.
We solved the Cahn-Hilliard equation numerically on a square domain and demonstrated the configuration of patch formations in the parameter space.
Next, we simulated the fully coupled system in the case of small deformations from a flat plane in Section~\ref{sec:Monge} and studied the effect of bending energy on the dynamics of aggregation and diffusion of proteins.
Our results show that coupling between curvature, protein aggregation, and diffusion can lead to a strong mechanical feedback loop stabilizing the protein microdomains in regions of high curvature.

\section{Model development \label{sec:model}}

{ We first formulate the governing equations for coupled diffusion and aggregation of curvature-inducing proteins on a deformable viscous lipid membrane with bending elasticity, building on previous models \cite{mahapatra2020transport,rangamani2013interaction,Agrawal2011-qi}. 
We begin by formulating a free energy function for the membrane and apply the principle of energy minimization to derive the governing equations.
Complete details of the derivation are provided in the Electronic Supplementary Information (ESI). 

\subsection{Free energy of the membrane}
Our system consists of the lipids that comprise the membrane and transmembrane proteins that are embedded in the plane of the membrane and are capable of inducing curvature (\Cref{fig:schematic}).
Our model does not include the binding or unbinding of proteins from the bulk or the interactions of the bulk fluid with the membrane. 
The lipid bilayer is modeled as a thin elastic shell with negligible thickness that can bend out of the plane and be subject to in-plane viscous flow. 
Importantly, we assume that the membrane is areally incompressible and this constraint is imposed on the membrane using a Lagrange multiplier. 
Additionally, we use a continuum description for the protein distribution on the membrane.
We describe the different energy contributions to the total energy of the system in detail below.

\subsubsection{Protein diffusion}
The diffusion of proteins on the membrane surface is modeled using the principle of entropy maximization \cite{safran2018statistical}.
The entropy $S$ of $q$ proteins on $n$ binding sites can be found from the number of combinations, $\Omega=\,^{n\!}C_q$, and is given by
\begin{equation}
    S=k_B \log \Omega,
\end{equation}
where $k_B$ is the Boltzmann constant \cite{laurendeau2005statistical}.
For sufficiently large values of $q$ and $n$, {the entropic component} of energy per binding site can be represented as a function of area fraction  $\phi=q/n$ as  \cite{safran2018statistical},
\begin{equation}
\label{eqn:entropy}
\frac{W_{entropy}}{n}=-\frac{TS}{n}=k_{B} T[\phi \log \phi+(1-\phi) \log (1-\phi)],
\end{equation}
where $T$ is the temperature of system. 
Note that the area fraction $\phi$ can {also} be represented as the ratio of the local protein density, $\sigma$, and the saturation density of proteins on the surface, $\sigma_s$.
The energy density per unit area of the membrane is obtained by multiplying the energy density per binding site ($W_{entropy}/n$) with the saturation density of the proteins ($\sigma_s$). 
Note that the entropic component of the energy $W_{entropy}$ {is minimized} when the entropy $S$ is maximum, which corresponds to a uniform distribution of the proteins in the domain.
\subsubsection{Protein aggregation}

Aggregation of proteins, on the other hand, can be modeled using the interaction enthalpy of particles in a binary system. 
With the help of mean-field theory, a continuum representation of the aggregation energy per binding site can be derived as \cite{safran2018statistical,veksler2007phase,givli2012stability}
\begin{equation}
\label{eqn:aggregation}
\frac{W_{aggregation}}{n}=\frac{\gamma}{2} \phi(1-\phi)+ \frac{\gamma }{4 \sigma_s }|\nabla \phi|^{2},
\end{equation}
where $\gamma$ is the net effective interaction energy of the proteins. 
 This term captures protein-protein attraction when $\gamma>0$ and protein-protein repulsion when $\gamma<0$.

\subsubsection{Bending energy of the membrane}
We model the curvature elastic energy density of the membrane using the Helfrich Hamiltonian \cite{helfrich1973elastic} given by
\begin{equation}
\label{eqn:Helfrich}
W_{bending} =  \kappa[H-C(\sigma)]^2 + \bar{\kappa} K. 
\end{equation}
Here, $H$ and $K$ are mean and Gaussian curvatures of the membrane, $\kappa$ and $\bar{\kappa}$ are the bending and Gaussian rigidities, and $C$ is the spontaneous curvature induced by the proteins.
The spontaneous curvature is assumed to depend linearly on protein density~$\sigma$ \cite{Agrawal2011-qi,mahapatra2020transport} as
\begin{equation}
    C(\sigma)=\ell \sigma,
\end{equation}
where the proportionality constant, $\ell$, has units of length. 

We obtain the total energy density of the membrane, in terms of protein area fraction $\phi$, by combining  \Cref{eqn:entropy,eqn:aggregation,eqn:Helfrich} as
\begin{equation}
\begin{split}
\label{eqn:energy_density}
    W = \underbrace{ \vphantom{\bigg(}  k_B T \sigma_s \left[ \phi \log \phi +(1-\phi) \log \left(1-\phi\right)\right]}_{\text{entropic }}   + \underbrace{ \vphantom{\bigg(}  \frac{\gamma \sigma_s }{2} \phi (1 -\phi)   + \frac{\gamma}{4 } |{\nabla} \phi|^2}_{\text{aggregation}} +\underbrace{ \vphantom{\bigg(}   \kappa(H-\ell \sigma)^2 + \bar{\kappa} K.}_{\text{bending}} 
\end{split}
\end{equation}


\subsection{Equations of motion}
The lipid bilayer is modeled as a two-dimensional surface in a three-dimensional space (\Cref{fig:schematic}b). 
We refer the reader to \cite{Steigmann1999-su,rangamani2013interaction, mahapatra2020transport} for details of the derivation and briefly summarize the key steps here. 
The equations of motion are obtained from a local stress balance on the interface, which can be compactly stated as 
\begin{equation}
     \nabla  \cdot \boldsymbol{\Sigma}+p \boldsymbol{n} =\mathbf{0},
\end{equation}
where $\boldsymbol\Sigma$ is the stress tensor, $\nabla \cdot \boldsymbol{\Sigma}$ is the surface divergence of the stress, 
$p$ is the normal pressure acting on the surface, and $\boldsymbol{n}$ is the unit surface normal vector.
As a result, the local equilibrium of forces, in the tangential and normal directions, is given by \Cref{eqn:stress_bal} and \Cref{eqn:stress_bal_normal} in the ESI.}
{The incompressibility constraint on the surface results in the following form of the continuity equation \cite{Steigmann1999-su}
\begin{equation}
    \nabla  \cdot \boldsymbol{v}=2Hw,
\end{equation}
where $w$ is the normal velocity of the surface.

\subsection{Mass conservation of proteins}
Conservation of mass for the protein density $\sigma$ is given by
\begin{equation}
\frac{\partial \sigma}{\partial t}+\nabla \cdot \boldsymbol{m}=0, 
\end{equation}
where the flux is 
\begin{equation}
\label{eqn:flux_def}
    \boldsymbol{m}=  \boldsymbol{v} \sigma -\frac{1}{f} \phi \nabla \mu  .
\end{equation}
This flux has contributions from advection due to the in-plane velocity field $\boldsymbol{v}$ and from gradients in the protein chemical potential $\mu$. The constant $f$ denotes the thermodynamic drag coefficient of a protein and is related to its diffusivity $D$ by the Stokes-Einstein relation: $D=k_B T/f$. 

The chemical potential, $\mu$, is obtained as the variational derivative 
\begin{equation}
\label{eqn:chem_pot_def}
    \mu=\frac{\delta F}{\delta \phi},
\end{equation}
where $F$ is the total energy of the system of area $A$, given by, 
\begin{equation}
\label{eqn:total_energy}
    F=\int_{\omega} W(\phi, \nabla \phi)\, dA.
\end{equation}
Note that the energy density is a function of both the protein area fraction $\phi$ and its gradient $ \nabla \phi$.
Using the definition of the variational derivative, we get the expression of the chemical potential as:
\begin{equation}
\label{eqn:chem_pot_den}
        \mu=\frac{\delta F}{\delta \phi}= \frac{\partial W}{\partial \phi}- \nabla  {\cdot} \frac{\partial W}{\partial  \nabla \phi}.
\end{equation}
Using \Cref{eqn:energy_density} for $W$ yields 
\begin{equation}
\begin{split}
\mu= k_B T \sigma_s [ \log \phi  -\log(1-\phi)] -2\kappa\ell \sigma_s (H-\ell \sigma_s \phi) 
-\frac{\gamma \sigma_s}{2} (2 \phi-1)- \frac{\gamma}{2} |\nabla \phi|^2.
\end{split}
\label{eq:potential}
\end{equation}
Substituting \Cref{eq:potential} in \Cref{eqn:flux_def} will result in the evolution equation for $\sigma$.

\subsection{System of governing equations}
Here we summarize the governing equations for the coupled dynamics of the system.
Using \Cref{eqn:traction,eqn:stress,eqn:vis_stress} for the stresses, 
the tangential force balance in \Cref{eqn:stress_bal} becomes \cite{Steigmann1999-su,rangamani2013interaction,mahapatra2020transport}
\begin{equation}
\label{eqn:tang_force_bal} 
\begin{split}
    \nabla \lambda &+ \underbrace{  2\nu  (\nabla \cdot \boldsymbol{d}-\nabla w \cdot \boldsymbol{b}) -4\nu w  \nabla  H }_{\text{viscous }} 
= \\
& \quad \quad \quad  \quad \quad -\nabla \sigma \bigg[ \underbrace{ \vphantom{\bigg(} k_B T \log  \frac{\phi}{1-\phi}}_{\text{entropic }} - \underbrace{ \vphantom{\bigg(}2\kappa\ell (H-\ell \sigma_s \phi) }_{\text{bending }} -\underbrace{\bigg(\frac{\gamma}{2} (2 \phi-1)+ \frac{\gamma}{2 \sigma_s} \Delta  \phi \bigg) }_{\text{aggregation }}\bigg].
\end{split}
\end{equation}
Here, we have introduced a new variable $\lambda$, which is the Lagrange multiplier for area incompressibility and physically represents the membrane tension (see \Cref{eqn:tension_lag} in the ESI for details), $\boldsymbol{d}$ is the rate-of-strain tensor (see \Cref{eqn:strain_rate} in the ESI for the details), $\boldsymbol{b}$ is the curvature tensor of the surface,
and $\Delta (\cdot)=\nabla  \cdot \nabla  (\cdot)$ is the surface Laplacian. 
Along with
the surface incompressibility condition
\begin{equation}
\label{eqn:continuity}
\nabla \cdot \boldsymbol{v}=2wH,
\end{equation}
 \Cref{eqn:tang_force_bal} describes how the surface pressure gradient is balanced by the tangential contributions of lipid flow, membrane bending, and membrane-protein interactions.
On the other hand, \Cref{eqn:continuity} captures surface incompressibility for a deformed membrane. 
\Cref{eqn:tang_force_bal,eqn:continuity} constitute the governing equations for the velocity field and tension on the evolving surface of the membrane.  

The shape of the surface is obtained by the normal force balance \Cref{eqn:stress_bal_normal}, which, after substituting in \Cref{eqn:stress}, \Cref{eqn:traction,eqn:vis_stress}, is given by
\begin{equation}
\label{eqn:norm_force_bal}
\begin{split}
     & \underbrace{\kappa\Delta (H- \ell\sigma_s \phi) +2\kappa(H-\ell\sigma_s \phi)(2H^{2}-K) -2H(\kappa(H-\ell \sigma_s \phi)^2 + \bar{\kappa} K)}_{\text{bending }} -\underbrace{ 2\nu \big[ \boldsymbol{b} : \boldsymbol{d} -w(4H^{2}-2K)\big]}_{\text{viscous }} \\
     & \quad \quad \quad \quad  -2H\Bigg[ \underbrace{ \vphantom{\bigg(} k_B T \sigma_s \{ \phi \log \phi 
     +(1-\phi) \log \left(1-\phi\right)\}}_{\text{entropic }}    +\underbrace{ \vphantom{\bigg(} \frac{\gamma \sigma_s }{2} \phi (1 -\phi) 
 + \frac{\gamma}{4  } |\nabla  \phi|^2}_{\text{aggregation }} \Bigg] =\underbrace{ \vphantom{\bigg(}  p + 2\lambda H.}_{\text{capillary}}
\end{split}
\end{equation}
While this equation is complex and contains many terms, it can be understood intuitively by making the following observations. 
In the absence of all other stresses (bending, viscous, entropic, and aggregation), \Cref{eqn:norm_force_bal} simply reduces to the Young-Laplace law.
When the viscous, entropic, and aggregation terms are removed, we recover the so-called `shape equation' that is commonly used in membrane mechanics \cite{Steigmann1999-su}.
The additional terms capture the non-trivial coupling between protein density, aggregation, lipid flow, and membrane bending, and are the novel aspect of the present model.
\Cref{eqn:tang_force_bal} and \Cref{eqn:norm_force_bal} both involve the area fraction of proteins $\phi=\sigma/\sigma_s$, which evolves according to the mass conservation equation given by
\begin{equation}
\begin{split}
\label{eqn:diff_agg_comp}
    \phi_t +  \nabla \cdot (\boldsymbol{v} \phi)=  & \,\frac{1}{f}\Delta \phi \left[   \frac{k_B T }{1-\phi} + {2\kappa \ell ^2 \sigma_s } \phi - { \gamma } \phi \right]-  \frac{1}{f}\phi \left[ {2\kappa \ell   } \Delta H  + \frac{\gamma}{2 \sigma_s} \Delta^2 \phi\right]  \\
    &  \quad \quad \quad \quad \quad \quad + \frac{1}{f} \nabla  \phi \cdot  \bigg[  \nabla \phi \bigg(    \frac{k_B T }{(1-\phi)^2} + {2\kappa \ell ^2 \sigma_s } -{\gamma } \bigg)  - {2\kappa \ell  }  \nabla H -\frac{\gamma}{2 \sigma_s}  \nabla  (\Delta \phi) \bigg].  
\end{split}
\end{equation}
Note that, in the absence of flow and protein-induced spontaneous curvature, \Cref{eqn:diff_agg_comp} reduces to the Cahn-Hilliard equation for aggregation-diffusion as discussed in Section~\ref{sec:CH}.
Additionally, if we eliminate protein aggregation (${\gamma}=0$), in the limit of dilute concentration of proteins ($\phi\ll 1$), we recover the classical equation for Fickian diffusion.





\subsection{Non-dimensionalization}
We non-dimensionalize the system of Equations (\ref{eqn:tang_force_bal})--(\ref{eqn:diff_agg_comp}) using the following reference scales. 
The characteristic length scale is taken to be the size $L$ of the domain.  The membrane tension $\lambda$ is scaled by its mean value $\lambda_0$. 
Velocities are non-dimensionalized by $v_c=\lambda_0 L/\nu$, and we use the diffusive time scale $t_c=L^2/D$. Note that the protein area fraction $\phi=\sigma/\sigma_s$ is already dimensionless. 
The governing equations in dimensionless form (where tildes are used to denote the dimensionless variables) are provided in the ESI (\Cref{eqn:tang_force_bal_nd}--\Cref{eqn:agg_diff_comp_nd}).
}
{The system of dimensionless equations involves seven dimensionless groups that are defined in \Cref{tab:dimensionless_numbers} along with their physical interpretation.
In all the analyses that follow, we assume that the transmembrane pressure, $p$, is zero. From here on, we use the dimensionless variables but omit the tildes for brevity. 
}
\begin{table}
\caption{List of dimensionless numbers and their definitions.}
\centering
\begin{tabular}{| c | c | c |}
\hline & & \\ [-1.0ex]
 Dimensionless Number & Expression & Physical interpretation \\  [1.0ex]
 \hline
 \hline  & & \\ [-1.5ex]
 $\hat{B}$ & $\displaystyle \frac{k_BT}{\kappa}$ & $\frac{\text{Thermal energy}}{\text{Bending energy}}$  \\  [2ex]
 $\hat{L}$ & $\displaystyle\frac{\ell}{L}$ &  $\frac{\text{Spontaneous curvature length}}{\text{Domain length}}$ \\ [2.5ex]
 $ \hat{A}$ & $\displaystyle\frac{\gamma}{k_B T}$ &  $\frac{\text{Aggregation coefficient}}{\text{Diffusion coefficient}}$ \\ [2.5ex]
 $\hat{S}$ & $\displaystyle\sigma_s L^2$ &  $\frac{\text{Domain area}}{\text{Protein footprint}}$ \\ [2ex]
 $\hat{T}$ & $\displaystyle\frac{2 L^2 \lambda_0}{\kappa}$ &  $\frac{\text{Membrane tension energy}}{\text{Bending energy}}$\\ [2ex]
 $Pe$ & $\displaystyle\frac{\lambda_0 L^2}{\nu D}$  &  $\frac{\text{Advection strength }}{\text{Diffusion strength}}$ \\ [1.5ex]
 \hline
\end{tabular}
\label{tab:dimensionless_numbers}
\end{table}

{
\subsection{Estimation of physical parameters}
Given the vast number of physical parameters in the model, we used data from the literature to estimate the ranges for these parameters and use these to inform the range of the dimensionless parameters in our simulations. 
We set the value of the bending rigidity $\kappa$  to 84 pN$\cdot$nm
\cite{schneider1984thermal,fowler2016membrane}. 
The range of spontaneous curvature length $\ell$ was chosen as 1-8 nm based on known protein-induced spontaneous curvature values \cite{quemeneur2014shape}.
The saturation density of proteins $\sigma_s$ on the lipid bilayer is varied in the range of $2\times10^{-4}$ to $2 \times 10^{-3}$ nm$^{-2}$, which corresponds to $20-70$ nm for the protein size \cite{horner2009coupled}. 
The viscosity of the membrane, $\nu$, is taken as 5 $\times 10^{-6}$ pN$\cdot$s/nm \cite{dimova1999falling,zgorski2019surface}, and the diffusion coefficient of a protein, $D$, is taken to be  5 $\times 10^{5}$ nm$^2$/s \cite{kahya2003probing,scherfeld2003lipid,kahya2004diff}.
For all the simulations, the domain is fixed as a square of side $L$ of 1 $\mu$m. 
The average membrane tension $\lambda_0$ is considered as 1 $\times 10^{-4}$ pN/nm \cite{baumgart2003Nature}. 
As a result, the P\'eclet number $Pe$ is fixed at 40, the range of $\hat{S}$ is 200 to 2000, and the range of $\hat{L}$ becomes 1 $\times 10^{3}$ to 8 $\times 10^{3}$. 
We found that the minimum value of $\hat{A}$ to promote aggregation is 11.1 based on stability analysis (Section~\ref{sec:CH_stability}), and considered the value of $\hat{A}$ in the range of 25 to 100 for the Cahn-Hilliard system (Section~\ref{sec:CH_num_sim}).
Although, for the coupled system of aggregation with bending, we used the value of $\hat{A}$ as 25 (Section~\ref{sec:LM_num_sim}).
The value of $\hat{B}$ at room temperature becomes $4.93 \times 10^{-2}$.
However, to demonstrate the dynamic coupling of aggregation and bending, we used a lower value of temperature $T$ in the numerical simulations, corresponding $\hat{B}$ was $4.93 \times 10^{-4}$.
This regime leads to a strong interaction between the membrane deformations and aggregation diffusion dynamics.
}

\label{sec:properties}
   

\section{Cahn-Hilliard system and stability analysis \label{sec:CH}}

\subsection{Reduction to the Cahn-Hilliard system}

We first consider the simplified diffusion-aggregation system in the absence of membrane bending and in-plane lipid flow to gain insight into how diffusion and aggregation compete in the plane of the membrane to form protein aggregates (also referred to as patterns or microdomains).
{We assume that the proteins have zero spontaneous curvature ($\hat{L}=0$) in this case.} 
As a result of these simplifications, the surface gradient 
reduces to the planar gradient ${\nabla =\frac{\partial }{\partial x}\boldsymbol{i}+\frac{\partial }{\partial y}\boldsymbol{j}}$ and the surface Laplacian  $\Delta $ becomes ${\nabla^2= \frac{\partial^2}{\partial x^2}+\frac{\partial^2}{\partial y^2}}$. 
Neglecting the flow and bending terms in \Cref{eqn:diff_agg_comp}, we arrive at a transport equation similar to the Cahn-Hilliard equation:
\begin{equation}
\begin{split}
\label{eqn:cahn_hilliard_nd}
     \phi_t  &   =  \nabla^2 \phi \bigg[  \frac{1}{1-\phi}   - \hat{A} \phi \bigg] + |\nabla  \phi|^2 \bigg[   \frac{1}{(1-\phi)^2}   -\hat{A} \bigg]    -  \phi \bigg[  \frac{\hat{A}}{2 \hat{S}} \nabla^4 \phi\bigg].
\end{split}
\end{equation}
\Cref{eqn:cahn_hilliard_nd} reduces to Fickian diffusion in the dilute limit ($\phi\ll 1$) in the absence of aggregation ($\hat{A}=0$).
\Cref{eqn:cahn_hilliard_nd} is also similar to the system presented by Givli and Bhattyacharya \cite{givli2012stability}, for which they conducted a stability analysis on a closed surface.
Here, we present a stability analysis of the equivalent Cahn-Hilliard system on a flat surface, and complement the analysis with numerical simulations of the nonlinear system in a periodic domain. 

\subsection{Linear stability analysis}
\label{sec:CH_stability}
\begin{figure}
    \centering
    \includegraphics[height=6cm]{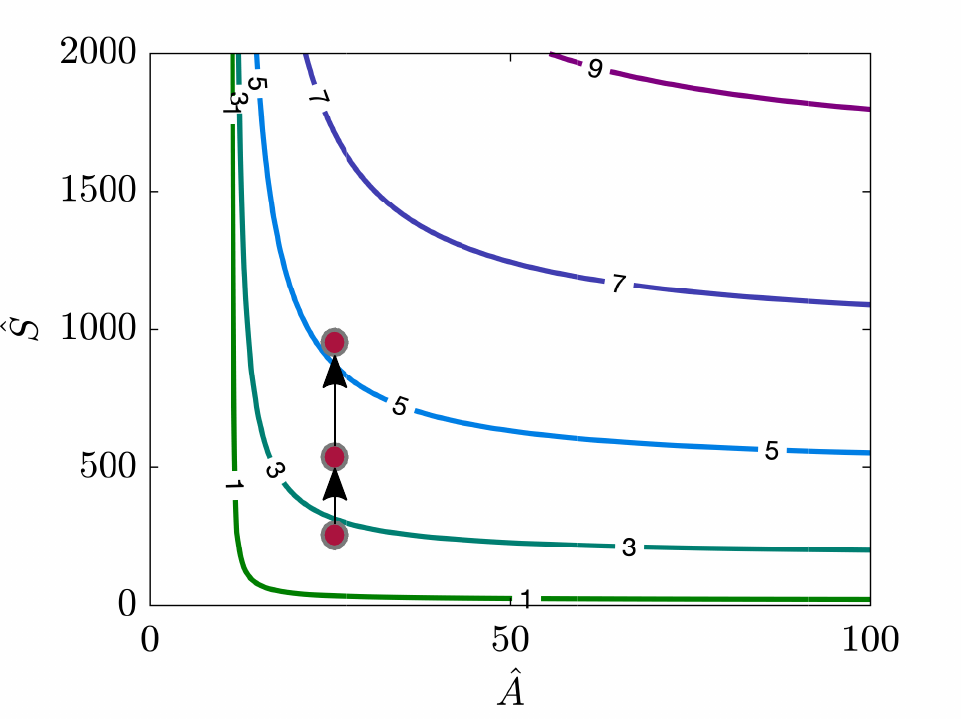}
    \caption{Marginal stability curves for the Cahn-Hilliard system in the ($\hat{A}$,$\hat{S}$) plane for $\phi_0=0.1$ and various wavenumbers $k$, as predicted by Equation \ref{eqn:cahn_hilliard_nd}. We mark three points in this figure to identify the parameter values for which we perform nonlinear numerical simulations in Figure~\ref{fig:CH_DA_dyn}.}
    \label{fig:dispersion}
\end{figure}

\begin{figure}
    \centering
    \includegraphics[width=0.95\textwidth]{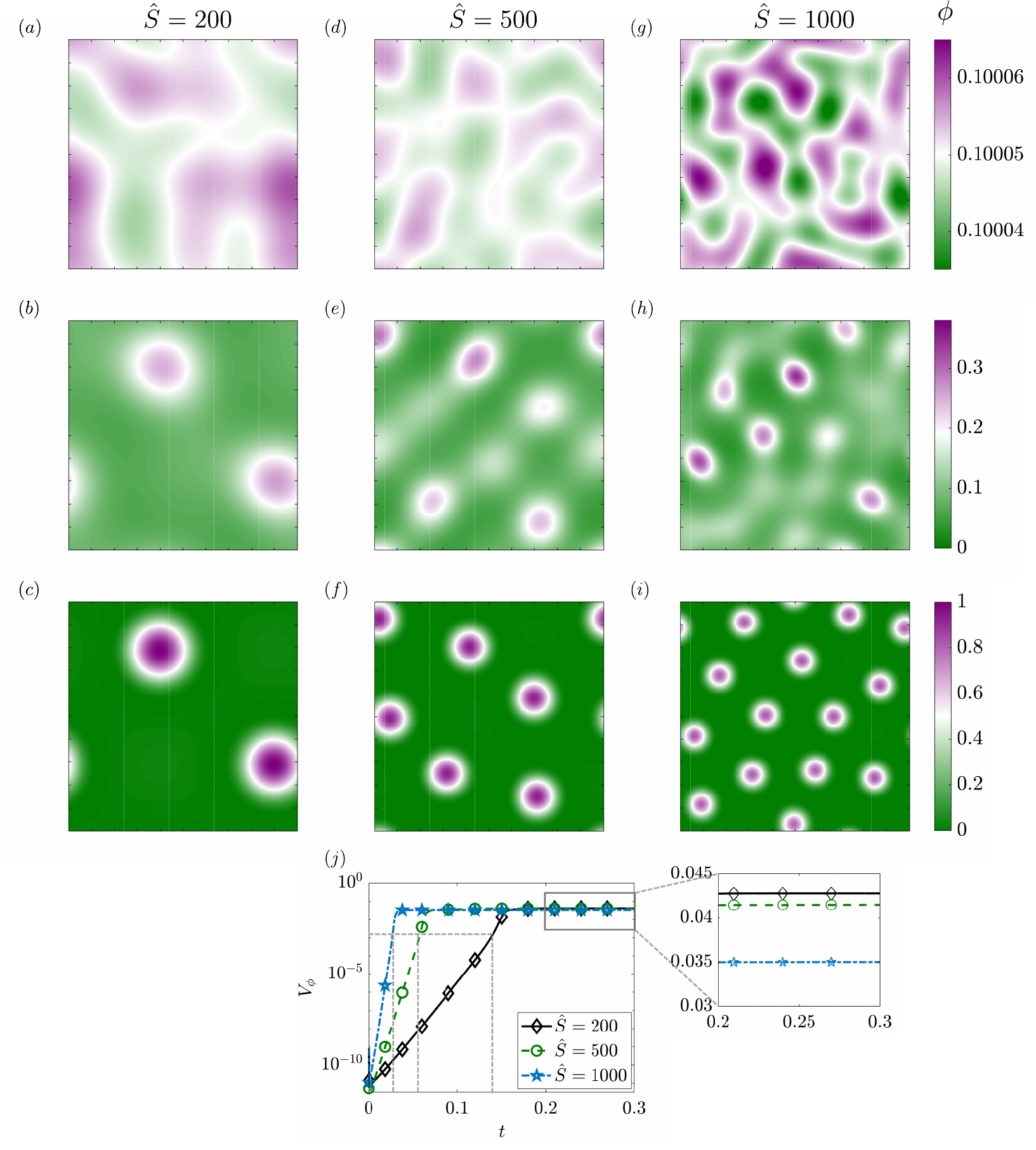}
    \caption{Temporal evolution of the protein distribution in simulations of the Cahn-Hilliard model of Figure \ref{eqn:cahn_hilliard_nd} on a flat square patch  of area 1 $\mu$m$^2$ for $\hat{A}=25$ and three different values of $\hat{S}$. 
    The three rows in panels (\textit{a-i}) correspond to three distinct times: at an early time $t_b=3\times 10^{-3}$ shortly after the start of the simulation, at an intermediate time $t_{in}$ when protein density variance reaches $V_{\phi}=2 \times 10^{-3}$, and at a late time $t_s=0.3$ when the system has reached steady state. The three columns correspond to $\hat{S}=200$
    (\textit{a-c}), $\hat{S}=500$
    (\textit{d-f}), and $\hat{S}=1000$ (\textit{g-i}).
    Also see Movies M1-M3 in the ESI for the corresponding dynamics.
    (\textit{j}) Temporal evolution of the variance $V_{\phi}$ of the protein density for the same cases shown in (\textit{a-i}). The dashed lines indicate the intermediate time $t_{in}$ when the variance reaches $V_{\phi}=2 \times 10^{-3}$. }
    \label{fig:CH_DA_dyn}
\end{figure}

We perform a linear stability analysis of \Cref{eqn:cahn_hilliard_nd} to identify the parameter regimes that can lead to protein aggregation.
The homogeneous state with uniform concentration $\phi_0$ is perturbed by a small amount $\phi^{\prime}$ such that $\phi=\phi_0 + \phi^{\prime}$.
Linearizing \Cref{eqn:cahn_hilliard_nd} results in the equation for density fluctuation $\phi^{\prime}$ as
\begin{equation}
\begin{split}
\label{eqn:linearize_CH}
    \phi^{\prime}_t   =  \nabla^2 \phi^{\prime} \bigg[  \frac{1}{1-\phi_0}   - \hat{A} \phi_0 \bigg] 
     -    \frac{\hat{A}}{2 \hat{S}} \phi_0 \nabla^4 \phi ^{\prime}. 
\end{split}
\end{equation}
We consider normal modes of the form $\phi^{\prime}= e^{\alpha t} e^{i 2\pi \boldsymbol{k}\cdot \boldsymbol{x}}$ and obtain the dispersion relation
\begin{equation}
    \alpha= 4 \pi^2\bigg[ \hat{A} \phi_0- \frac{1}{1-\phi_0}  \bigg] k^2 - 8\pi^4 \frac{\hat{A}}{\hat{S}} \phi_0 k^4.
    \label{eq:dispersion}
\end{equation}
We find that the growth rate $\alpha$ is always real. The first term in \Cref{eq:dispersion} is positive and is destabilizing as long as {the strength of aggregation exceeds a certain threshold:} $\hat{A}\ge \hat{A}_c= [\phi_0(1-\phi_0)]^{-1}(\approx 11.1$ for $\phi_0=0.1$),
whereas the second term is always stabilizing. The marginal stability curves $\alpha=0$  in the $(\hat{A},\hat{S})$ plane are plotted for various wavenumbers $k$ in \Cref{fig:dispersion}. For a given choice of $\hat{A}$ and $\hat{S}$, this results in a band of unstable wavenumbers $0\le k \le k_c$, where 
\begin{equation}
\label{eqn:nec_cond_agg}
        k_c^2 =\frac{\hat{S}}{2 \pi^2} \bigg[1-\frac{\hat{A}_c}{ \hat{A}} \bigg], 
\end{equation}
and the maximum growth rate occurs at wavenumber $k_m=k_c/\sqrt{2}$. The corresponding wavelength $\Lambda=2\pi/k_m$ provides a prediction for the characteristic lengthscale of aggregation patches, which is expected to decay with increasing $\hat{S}$ but to increase with increasing $\hat{A}$.  


\subsection{Numerical simulations}
\label{sec:CH_num_sim}
We conducted numerical simulations of \Cref{eqn:cahn_hilliard_nd} inside a square domain for various combinations of $\hat{A}$ and $\hat{S}$
that satisfy the necessary condition of aggregation as given in  \Cref{eqn:nec_cond_agg} and \Cref{fig:dispersion}. 
The initial condition was set as a homogeneous distribution of $\phi_0=0.1$ with a small random spacial perturbation with magnitude $|\phi^{\prime}|\le 1 \times 10^{-4}$. We numerically restricted the value of $\phi$ to the interval $[\epsilon,1-\epsilon]$ with $\epsilon=1\times 10^{-3}$ to ensure that neither $\phi$ or $1-\phi$ becomes zero during the simulations. 
We used periodic boundary conditions for the protein density and solved the equation numerically using a finite difference technique (the Fortran code is available on \href{https://github.com/armahapa/protein_aggregation_in_membranes}{\Blue{https://github.com/armahapa/protein\_aggregation\_in\_membranes}}). 
In \Cref{fig:CH_DA_dyn}, we show the evolution of the protein distribution over time for three different values of the dimensionless number $\hat{S}$ {that denotes the ratio of domain area to the protein footprint}, while  {maintaining the aggregation strength at} $\hat{A}=25$. In all cases, we find that the initial perturbation in the density field evolves towards the formation of distinct dense circular protein patches that are distributed randomly and nearly uniformly across the domain, in agreement with standard Cahn-Hilliard aggregation dynamics \cite{gera2017cahn}. The main effect of varying $\hat{S}$, which is more dramatic than varying $\hat{A}$ as we further show below, is to control the number of patches as well as their size. Indeed, we recall that $\hat{S}$, which is a dimensionless measure of the finite size of the proteins, directly controls the stabilizing term in the dispersion relation \Cref{eq:dispersion} and therefore the dominant wavenumber of the instability. Consistent with the stability predictions, we find that larger values of $\hat{S}$ produce larger numbers of patches with smaller sizes. During the transient evolution, proteins get drawn towards the emerging patches due to aggregation, and at steady state we find that the density inside the patches approached the saturation density ($\phi=1$), whereas it approaches zero outside (Also see Movies M1-M3 in the ESI). We quantify the growth of density fluctuations by plotting in \Cref{fig:CH_DA_dyn}$j$ the time evolution of the density variance, defined as
\begin{equation}
    V_{\phi}=\int_A (\phi-\phi_0)^2 dA.
\end{equation}
We find that the growth of the variance is exponential at short times, consistent with the expected behavior for a linear instability, before reaching a constant plateau at long times. The growth is observed to increase with $\hat{S}$ in agreement with the linear prediction of \Cref{eq:dispersion}. The steady state value, on the other hand, is found to decrease slightly with $\hat{S}$, although the differences are small. 



\begin{figure*}
    \centering
    \includegraphics[width=0.95\textwidth]{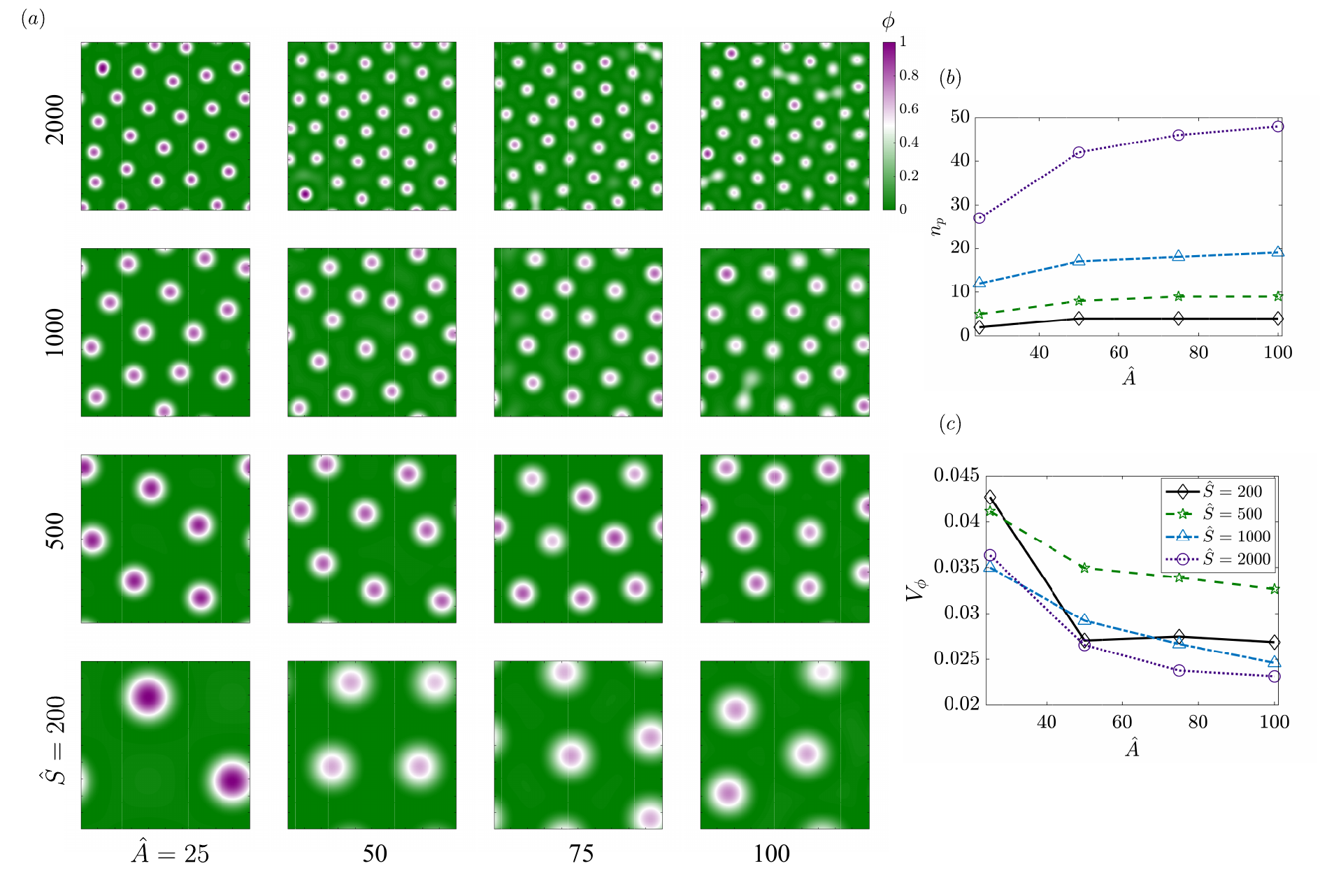}
    \caption{(\textit{a}) Configurations of protein aggregates on a flat square membrane at a late time $t=0.3$ approaching steady state for  various combinations of $\hat{A}$ and $\hat{S}$.  
    (\textit{b}) Variation of the number of protein patches with $\hat{A}$, for different values of $\hat{S}$. (\textit{c}) Variation of of the protein density variance $V_{\phi}$ with $\hat{A}$ for different values of $\hat{S}$. }
    \label{fig:CH_sig_phase}
\end{figure*}


A more complete exploration of pattern formation is provided in \Cref{fig:CH_sig_phase}\textit{a}, showing the long-time configurations of aggregated protein patches in the parameter space of $\hat{A}$ and $\hat{S}$. 
We note that the number of patches, their size, and their homogeneity vary with both parameters. As we already observed in \Cref{fig:CH_DA_dyn},  increasing $\hat{S}$ for a given value of $\hat{A}$ increases the number of patches and decreases their size. On the other hand, increasing $\hat{A}$ for a given $\hat{S}$ tends to increase inhomogeneity among patches, with some visibly denser patches while others tend to be more diffuse.
The dependence of the number of patches as a function of both $\hat{A}$ and $\hat{S}$ is shown in \Cref{fig:CH_sig_phase}\textit{b}, while the steady-state variance is plotted in \Cref{fig:CH_sig_phase}\textit{c}. The variance is found to decrease with $\hat{A}$, as the more  diffuse patches forming at large $\hat{A}$ result in weaker spatial fluctuations. 


\section{Coupling of aggregation with bending: analysis in the small deformation regime \label{sec:Monge}}

To understand how the inclusion of membrane curvature alters the aggregation-diffusion landscape, we simulated the dynamics of the coupled system \Cref{eqn:tang_force_bal_nd,eqn:cont_nd,eqn:norm_force_bal_nd,eqn:agg_diff_comp_nd} in the regime of small deformations from a plane.
The surface is represented using the Monge parametrization, such that the position vector is given by $\boldsymbol{r}=x_\alpha\boldsymbol{e}_\alpha+z(x_1, x_2, t)\boldsymbol{e}_3$.
In the regime of small deformations from the plane, we consider gradients of the surface deformation to be small and ignore the higher-order terms \cite{mahapatra2020transport}.
The surface gradient and Laplacian in the Monge parameterization simplify to
${\nabla}=\frac{\partial }{\partial x}\boldsymbol{i}+\frac{\partial }{\partial y}\boldsymbol{j}$ and $\nabla^2= \frac{\partial^2}{\partial x^2}+\frac{\partial^2}{\partial y^2}$. 
In the limit of small deformations, the system of governing equations \Cref{eqn:cont_nd} to \Cref{eqn:agg_diff_comp_nd} reduces to \Cref{eqn:LM_cont} to \Cref{eqn:LM_diff}.
 

\subsection{Linear stability analysis \label{sec:stability2}}
\begin{figure}[t]
    \centering
    \includegraphics[width=0.5\textwidth]{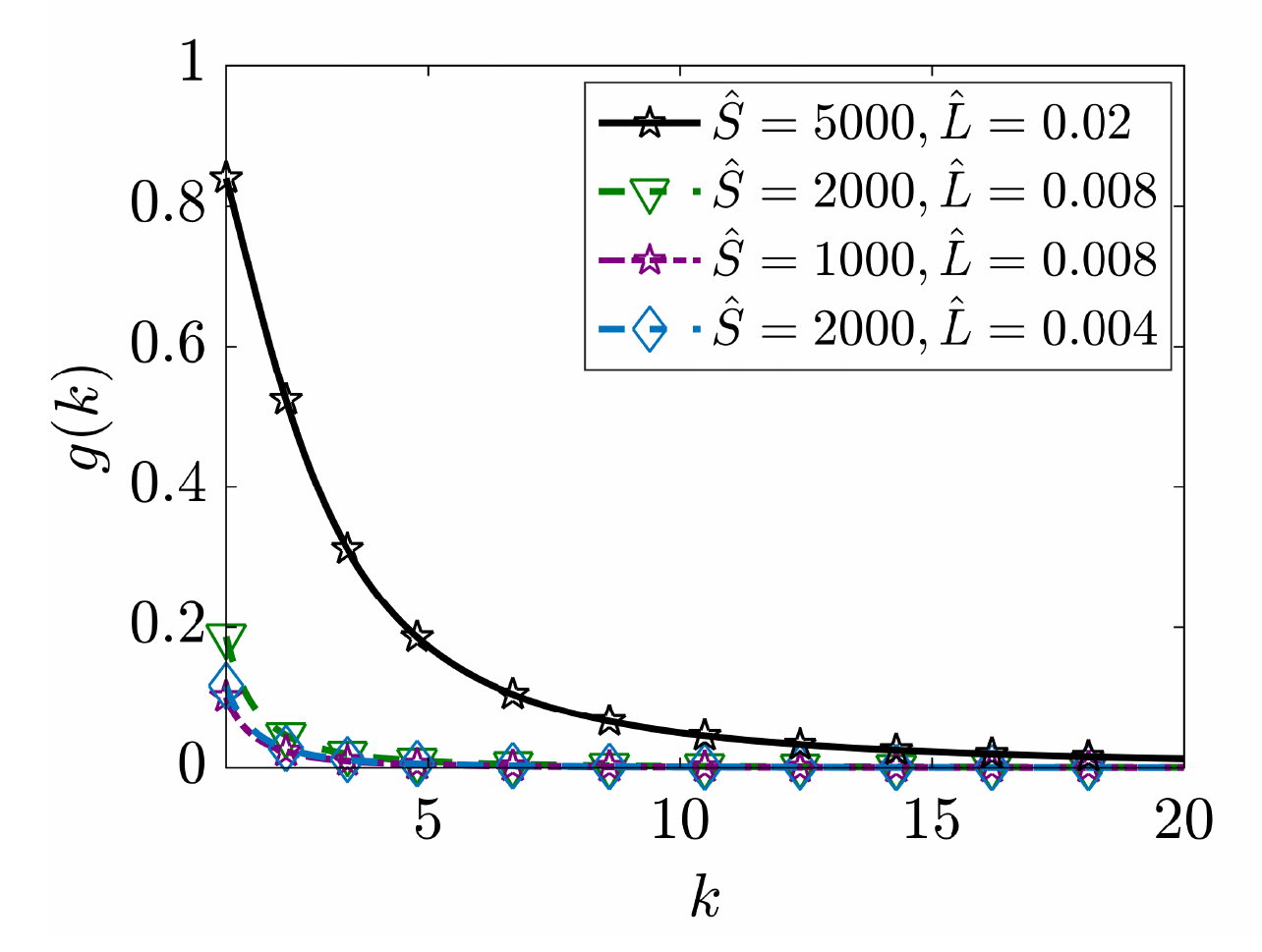}
    \caption{Dependence of $g$ defined in \Cref{eq:functiong} on wavenumber $k$ for different values of $\hat{S}$ and $\hat{L}$, with $\hat{A}=25$. }
    \label{fig:gk}
\end{figure}

We first perform a stability analysis of the system of equations (\Cref{eqn:LM_cont} to \Cref{eqn:LM_diff}) to identify the parameter regimes similar to the analysis of Section~\ref{sec:CH_stability} but in the presence of bending due to spontaneous curvature induced by the protein. In the base state, the membrane is flat and at rest with uniform tension ($z_0=0$, $\boldsymbol{v}_0=\boldsymbol{0}$, $\lambda_0=1$), and the protein density is uniform with value $\phi_0$. We showed in an earlier study \cite{mahapatra2020transport} that a uniform protein distribution on a flat membrane is indeed a steady state even when the proteins induce a spontaneous curvature. We perturb the variables by small amounts with respect to this base state:
\begin{equation}
    \phi=\phi_0 + \phi^{\prime}, \quad z=0+ z^{\prime},\quad \boldsymbol{v} = \boldsymbol{0}+\boldsymbol{v}^{\prime}, \quad 
    \text{and,} \quad \lambda  = 1+\lambda^{\prime}. 
\end{equation}
Linearizing \Cref{eqn:LM_cont,eqn:LM_tang} provides the governing equations for velocity and tension fluctuations as 
\begin{equation}
    {{\nabla}}\cdot \boldsymbol{v}^{\prime}=0,
       \label{eqn:lin_LM_cont}
\end{equation}
 and,
\begin{equation}
\begin{split}
   \quad {\nabla} {\lambda}^{\prime} + \nabla^2 {\boldsymbol{v}}^{\prime} + \nabla (\nabla \cdot {\boldsymbol{v}}^{\prime})=-{\nabla} \phi ^{\prime} \bigg[  \frac{2 \hat{B} \hat{S}}{\hat{T}}   \log \frac{\phi_0}{1-\phi_0} +
 \frac{4 \hat{L}^2 \hat{S}^2}{\hat{T}} \phi_0-\frac{\hat{A} \hat{B} \hat{S}}{\hat{T}} (2 \phi_0 -1) \bigg].
   \label{eqn:lin_LM_tang}
\end{split}
\end{equation}
The normal force balance of \Cref{eqn:LM_norm} reduces to
\begin{equation}
\begin{split}
     \nabla^4 {z}^{\prime}-2\hat{L}\hat{S}\nabla^2 \phi ^{\prime} 
     -2  \hat{B} \hat{S}\nabla^2 {z}^{\prime} \bigg[\{\phi_0 \log \phi_0 +(1-\phi_0)\log (1-\phi_0)\} 
       +\frac{\hat{A}}{2} \phi_0 (1-\phi_0) 
      +\frac{\hat{L}^2 \hat{S} }{\hat{B}} \phi_0^2\bigg]=\hat{T} \nabla^2 {z}^{\prime}  .
       \label{eqn:LM_norm_lin}
\end{split}
\end{equation}
Finally, the transport equation for the protein density given in \Cref{eqn:LM_diff} becomes  
\begin{equation}
\begin{split}
    \phi^{\prime}_t  & =  \nabla^2 \phi^{\prime} \bigg[  \frac{1}{1-\phi_0} + \frac{2\hat{L}^2 \hat{S}}{\hat{B} } \phi_0 - \hat{A} \phi_0 \bigg]-  \phi_0 \bigg[\frac{\hat{L}}{ \hat{B}} \nabla^4 z^{\prime}  + \frac{\hat{A}}{2 \hat{S}} \nabla^4 \phi ^{\prime} \bigg].  \label{eqn:LM_diff_lin}
\end{split}
\end{equation}
We find that the linearized equations the velocity field and tension {partially decouple} from the shape equation (\ref{eqn:LM_norm_lin}) and protein transport equation (\ref{eqn:LM_diff_lin}): in other words, lipid flow and tension fluctuations do not affect the membrane shape and protein transport in the linear regime. 
To analyze the dynamics of protein aggregation, we therefore need only consider \Cref{eqn:LM_norm_lin,eqn:LM_diff_lin}. Performing a normal model analysis (see ESI for details), we obtain the dispersion relation as 
\begin{equation}
\label{eqn:LM_dispersion}
    \alpha   = 4\pi^2 k^2 \bigg[  \hat{A} \phi_0-   \frac{1}{1-\phi_0}  - \frac{2\hat{L}^2 \hat{S}}{\hat{B} }  \phi_0 g(k) \bigg]  - 8 \pi^4 \phi_0 \frac{\hat{A}}{ \hat{S}}  k^4,
\end{equation}
where $g(k)$ is given by
\begin{equation}
    g(k)= 1-\frac{16 \pi^4 k^4}{M(k)}, \label{eq:functiong}
\end{equation}
and,
\begin{equation}
    M(k)= 16 \pi^4 k^4 + 8 \pi^2k^2 \hat{B} \hat{S}  \bigg[\{\phi_0 \log \phi_0 +(1-\phi_0)\log (1-\phi_0)\} 
      +  \frac{\hat{A}}{2} \phi_0 (1-\phi_0) 
      +\frac{\hat{L}^2 \hat{S} }{\hat{B}} \phi_0^2\bigg] + 4 \pi^2 k^2 \hat{T}.
\end{equation}
Similar to \Cref{eq:dispersion}, the second term in \Cref{eqn:LM_dispersion} is always stabilizing, and therefore protein aggregation takes place only if the first term is positive.  
The necessary condition for protein aggregates to form becomes
\begin{equation}
    \hat{A} - \frac{2\hat{L}^2 \hat{S}}{\hat{B}} g(k) \ge \frac{1}{\phi_0 (1-\phi_0)},
\end{equation}
or 
\begin{equation}
\label{eqn:cond_LM_aggr}
    \hat{A} \ge \hat{A}_c+ \frac{2\hat{L}^2 \hat{S}}{\hat{B}}g(k),
\end{equation}
where $\hat{A}_c$ was previously defined in Section~\ref{sec:CH_stability} in the Cahn-Hilliard case.
{Here again, we find that there exists an unstable range of wave numbers $0<k<k_c$, where $k_c$ satisfies the implicit equation 
\begin{equation}
k_c^2=\frac{\hat{S}}{4 \pi^2} \bigg[1-\frac{\hat{A}_c}{A} -\frac{2 \hat{L}^2 \hat{S}}{\hat{B} \hat{A}} g(k_c)  \bigg].
\end{equation}
The maximum growth rate occurs at wavenumber $k_m$, also given by an implicit equation:}
\begin{equation}
{k_m = \frac{k_c}{\sqrt{2}}  \bigg[ 1+ \frac{1}{4 \pi^2 k_m} \frac{{\hat{L}}^2 {\hat{S}}^2}{\hat{B} \hat{A}} g^{\prime}(k_m) \bigg] ^{-1/2}.}
\end{equation}
Figure \ref{fig:gk} shows the dependence of $g(k)$ on wave number $k$ for $\hat{A}=25$ and various combinations of $\hat{L}$ and $\hat{S}$.
When both $\hat{L}$ and $\hat{S}$ increase, $g(k)$ tends to increase for small wavenumbers and thus stabilizes the system. This means in particular that proteins with large spontaneous curvature, as captured by $\hat{L}$, can in fact have a stabilizing effect on protein aggregation, and this counterintuitive observation will be confirmed in numerical simulations as we discuss next.

\subsection{Numerical simulations}
\label{sec:LM_num_sim}

\begin{figure*}
    \centering
    \includegraphics[width=0.95\textwidth]{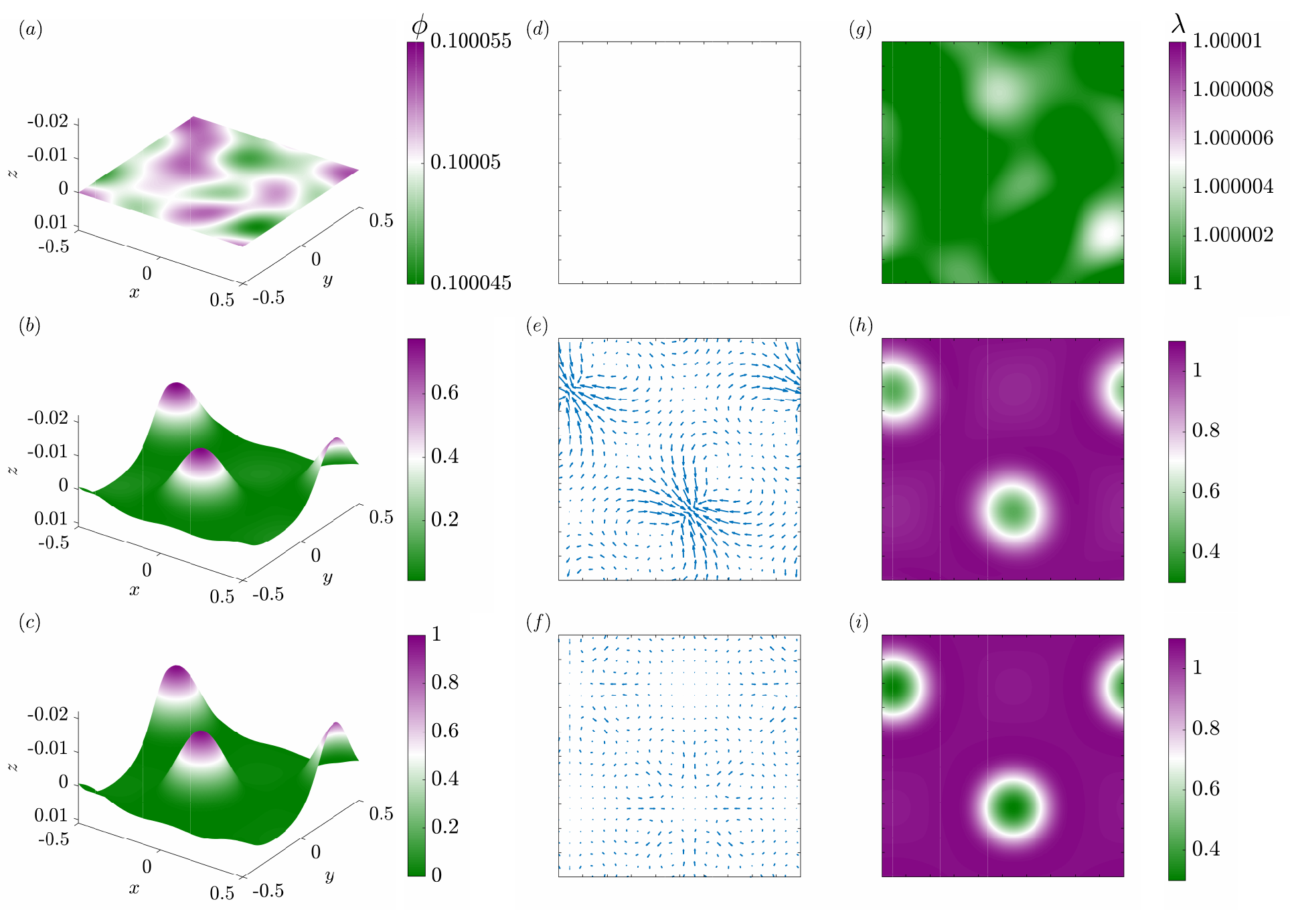}
    \caption{Temporal evolution of protein distribution, membrane shape, in-plane velocity and tension for a square membrane of size 1 $\mu$m$^2$ with $\hat{A}=25$, $\hat{S}=200$, and $\hat{L}=8 \times10^{-3}$. 
    (\textit{a-c}) Height of the membrane colored with the local protein density, 
    (\textit{d-f}) in-plane velocity field, and 
    (\textit{g-i}) membrane tension at dimensionless times $0.003$, $0.216$, and $0.3$. } 
    \label{fig:LM_AG_dyn}
\end{figure*}

\begin{figure*}
    \centering
    \includegraphics[width=\textwidth]{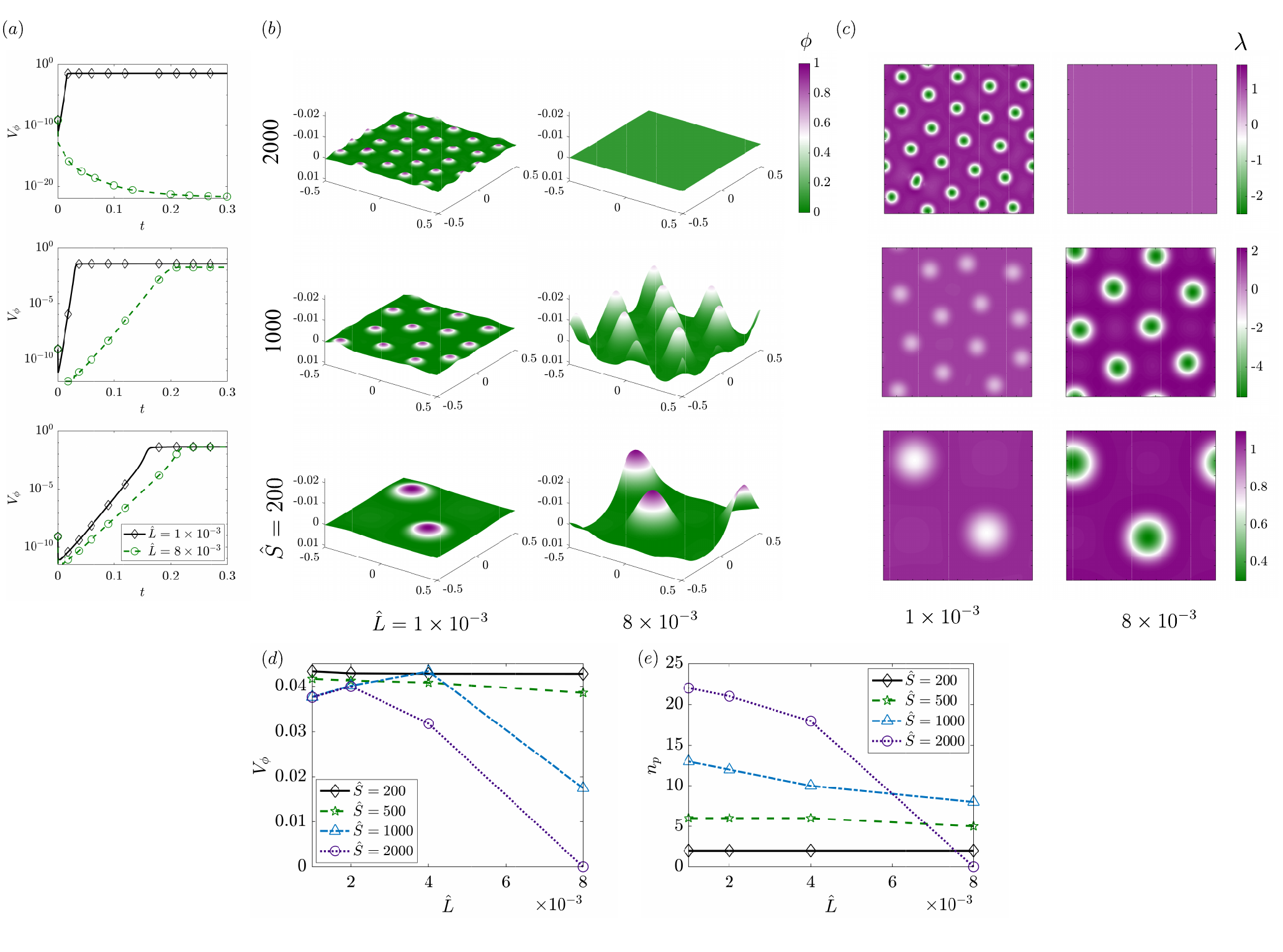}
    \caption{Effect of $\hat{S}$ and $\hat{L}$ on protein aggregation and membrane dynamics. (\textit{a}) Temporal evolution of the protein density variance $V_{\phi}$ for two values $\hat{L}$ and the same three values of $\hat{S}$ shown in (\textit{b}).  (\textit{b}) Distribution of protein density on the deformed membrane at a long time approaching steady state (${t}=0.3$) for various combinations of  $\hat{L}$ and $\hat{S}$, with $\hat{A}=25$.
    The corresponding dynamics are also shown in movies M4-M6 of the ESI.
    (\textit{c}) Distribution of the local membrane tension for the same cases as in ($b$).
    (\textit{d}) Variance of protein density $V_{\phi}$ and ($e$) number of protein patches $n_p$ at ${t}=0.3$ as functions of $\hat{L}$, for various values of $\hat{S}$.      
    }
    \label{fig:patch_con_LM}
\end{figure*}

We solved \Cref{eqn:LM_cont,eqn:LM_tang,eqn:LM_norm,eqn:LM_diff} numerically on a square domain with periodic boundary conditions for a small random density perturbation over a homogeneous steady state density of $\phi=0.1$. The proteins now induce a spontaneous curvature in the membrane, and the model also captures the viscous flow on the membrane manifold. Typical transient dynamics are illustrated in \Cref{fig:LM_AG_dyn} in a simulation with $\hat{L} = 8 \times 10^{-3}$, $\hat{A}=25$, and $\hat{S}=2000$. 
The initial random distribution resolves into strong patches of proteins over time with the same number of patches as we observed in the Cahn-Hilliard system (compare \Cref{fig:CH_DA_dyn}\textit{a-c} with \Cref{fig:LM_AG_dyn}\textit{a-c}).
Because the system of equations now accounts for coupling of curvature with protein dynamics, we observe that the formation of dense protein patches is accompanied by the localized growth of membrane deformations, in the form of nearly circular peaks surrounded by flatter regions of oppositely-signed curvature (\Cref{fig:LM_AG_dyn}\textit{a-c}). 
We also observe that the formation of protein aggregates is coupled with a tangential velocity field in the plane of the membrane, to accommodate the deformation of the membrane  (\Cref{fig:LM_AG_dyn} \textit{d-f}): as the protein aggregates form and deflect the membrane in the normal direction, a source-like flow is generated locally as dictated by the continuity relation \Cref{eqn:LM_cont}.
During this process, the magnitude of the velocity increases until the system approaches a steady state where aggregation balances diffusion. As the steady state is approached, the flow in the membrane changes nature as the normal velocity vanishes, with each protein patch driving a weaker flow with quadrupolar symmetry. 

As we have noted in prior works \cite{rangamani2013interaction,rangamani2014,mahapatra2020transport}, coupling of lipid flow to membrane deformation not only completes the description of the physics underlying the viscoelastic nature of the membrane but also allows for the accurate calculation of the membrane tension field (the Lagrange multiplier for incompressibility).
This is particularly relevant for understanding how microdomains of proteins can alter the tension field in the membrane.
The tension field on the membrane tracks with the protein microdomains and the deformation in the coupled system (\Cref{fig:LM_AG_dyn}\textit{g-i}). 
Initially, the membrane has nearly uniform tension, but as regions of high protein aggregation and therefore high membrane curvature form, these locations are found to have lower tension in comparison with the rest of the membrane (see \cite{rangamani2014} for a detailed discussion on this point). 
Thus, the dynamics of the coupled system is able to capture key experimental observations in the field of membrane-protein interactions: (a) regions of high curvature and aggregation are correlated for curvature-inducing proteins suggesting a positive feedback between these two important factors \cite{baumgart2007PNAS}, (b) lipid flow is important to sustain the deformations (\cite{saeki2006growth}), and (c) membrane tension is a heterogeneous field and varies with the local membrane composition \cite{baumgart2003Nature}. 

To further quantify these behaviors, we investigated the parameter space of $\hat{S}$ and $\hat{L}$, to understand how the spontaneous-curvature induction versus protein footprint compete in a fixed regime of aggregation-to-diffusion ($\hat{A}=25$ fixed) (see \Cref{eqn:cond_LM_aggr}). 
We varied $\hat{S}$ in the range of $200$ to $2000$ and $\hat{L}$ from $1 \times 10^{-3}$ to $8\times 10^{-3}$ and summarize these results in \Cref{fig:patch_con_LM}.
We first observed that the growth rate of the variance of $\phi$ shows a strong dependence on $\hat{L}$ (\Cref{fig:patch_con_LM}\textit{a}). 
For $\hat{S}=200$, the growth rate for the two different values of $\hat{L}$ differ slightly with the growth rate being slower for larger $\hat{L}$.
This effect persists and is amplified for larger $\hat{S}$: as both $\hat{S}$ and $\hat{L}$ increase, the growth rate decreases, indicating that it takes longer time for patterns to form on the membrane. 
However, when $\hat{S}=2000$, we see a decay in the variance of protein density $\phi$ as opposed to the exponential growth and eventual plateau for the cases where protein aggregrates form. 
This result, which is consistent with the stability analysis of Section~\ref{sec:stability2} suggests that the induction of curvature on the membrane can alter significantly the dynamics of protein aggregation.

The steady-state patterns and deformations are illustrated in \Cref{fig:patch_con_LM}\textit{b} (also see Movies M4-M6, and \Cref{fig:LM_z_appendix,fig:LM_Lam_appendix} in the ESI), where we observe that the number of protein patches is largely unaffected by $\hat{L}$ for  $\hat{S}=200$.
The number of patches increases with $\hat{S}$ for a given $\hat{L}$ (as already found in \Cref{fig:CH_sig_phase}).
However, when $\hat{S}$ increases to $1000$, the number of patches decrease with $\hat{L}$.
Since the deformation is directly affected by spontaneous curvature, we find however that $\hat{L}$ has a strong effect on the magnitude of membrane deflections, with larger protein footprints resulting in stronger deflections. 
Surprisingly, when  $\hat{S}=2000$, we noticed that protein aggregates do not form for the value of $\hat{L}=8\times 10^{-3}$ and the membrane remains flat.  This phenomenon can be explained from the critical value of $\hat{A}$ in \Cref{eqn:cond_LM_aggr}. Since both $\hat{L}$ and $\hat{S}$ have a stabilizing effect on density fluctuations $\phi^{\prime}$ (\Cref{eqn:cond_LM_aggr}), for higher value of $\hat{S}$ and $\hat{L}$, an aggregation coefficient of $\hat{A}=25$ is not sufficient to overcome the stabilizing barrier. 
However, for lower values of $\hat{L}$ or lower values of $\hat{S}$, where the stabilizing effect is relatively weak, we see the formation of protein aggregates. 

The tension profile in the membrane follows the inhomogeneity of the protein distribution as expected (\Cref{fig:patch_con_LM}\textit{c} and \Cref{fig:LM_Lam_appendix}). As previously noted in \Cref{fig:LM_AG_dyn}, the patches are associated with tension minima. 
{We find that the range of $\lambda$ depends strongly on $\hat{S}$ and $\hat{L}$, as ${\nabla} \lambda$ linearly depends on the negative of the gradient of the spontaneous curvature, which in turn depends on both $\ell$ and $\sigma$. 
This is consistent with our previous results showing that $\lambda$ is a heterogeneous field on the membrane and varies with the protein-induced spontaneous curvature \cite{rangamani2014,mahapatra2020transport}.} 
\Cref{fig:patch_con_LM}$c$ further highlights the coupling between curvature, flow, and aggregation dynamics.
 Finally, we look at the variance and the number of patches as a function of both $\hat{S}$ and $\hat{L}$ (\Cref{fig:patch_con_LM}\textit{d,e}).
 We note that for a given value of $\hat{S}$, the variance decreases with increasing $\hat{L}$ for higher values of $\hat{S}$ and this decrease is more dramatic when compared to the Cahn-Hilliard model (\Cref{fig:CH_sig_phase}\textit{b}).
 Even though the number of patches remains more or less unaltered for small values of $\hat{S}$ as $\hat{L}$ increases, the number decreases with increasing $\hat{L}$ for larger values of $\hat{S}$ (\Cref{fig:patch_con_LM}\textit{e}), consistent with the stability behavior noted in \Cref{eqn:cond_LM_aggr}.
These results suggest that the landscape of protein inhomogeneity is not only governed by the $\hat{A}$-$\hat{S}$ space as is the case in the Cahn-Hilliard model; rather the curvature parameters, specifically $\hat{L}$ in this case, can have a significant impact on the protein aggregation behavior.
Thus, we find that the aggregation-diffusion landscape on the surface of the membrane is altered by the protein-induced spontaneous curvature -- tuning these different effects can allow for differential control of curvature-aggregation feedback.

\section{Discussion}
The interaction of peripheral and integral membrane proteins with the lipid bilayer of cellular membranes is fundamental to cellular function \cite{McMahon2015-fi,McMahon2005-fk,Kozlov2014-rb}.
In this work, we have developed a comprehensive modeling framework that couples the multiple effects that take place in such membrane-protein interactions: protein diffusion in the plane of the membrane, interaction between the proteins resulting in aggregation, lipid flow in the plane of the membrane, and out-of-plane curvature generation due to protein-induced spontaneous curvature. 
The resulting system of equations now completely describes the mechanics of a lipid membrane with a second species that can both diffuse and aggregate in the plane of the membrane.
We compared this system against a reduced system of Cahn-Hilliard equations to show how the coupling with membrane bending alters the system behavior using both linear stability analysis and numerical simulations.
In the absence of curvature coupling (the Cahn-Hilliard system), the dynamics of protein aggregation is driven by the competition between two key parameters, $\hat{S}$, representing the relative size of the protein footprint and $\hat{A}$, representing the relative strength of protein aggregation over diffusion.
In the presence of curvature coupling due to protein-induced spontaneous curvature, these dynamics are altered and depend strongly on the strength of the spontaneous curvature induced by these proteins.
These altered dynamics can be summarized as follows: for certain regimes of $\hat{S}$ and $\hat{L}$, microdomains of proteins form on the membrane and are closely tied to the membrane curvature as is expected, generating a strong feedback between curvature and aggregation. 
We also found that for certain regimes of $\hat{S}$ and $\hat{L}$, the growth rate decays, preventing the formation of protein aggregates and the membrane remains flat. 

The interaction between curvature and protein aggregation in membranes has been studied in multiple modeling \cite{Agrawal2008-ru,Agrawal2008-rs,veksler2007phase,givli2012stability}, simulation \cite{reynwar2007aggregation,weber2019elife1,gera2017cahn}, and experimental contexts \cite{Callan-Jones2013-vn,Beber2019-ja,Aimon2014-ex,Stachowiak2012-th,Stachowiak2010-vy}.
Our work builds on this literature with a few key differences.
Many of the theoretical models analyze the governing equations in simplified settings.
In some cases, the geometry is fixed and the emergence of patterns is analyzed, and in other cases, the dynamics of the protein interactions on the surface is ignored \cite{givli2012stability,gera2017cahn}. 
Here, we have analyzed the fully coupled  system without any assumptions on the dominant regimes and demonstrated how curvature generation can affect aggregation.
Another important feature of our model is the calculation of membrane tension.
Since the lipid bilayer is assumed to be incompressible, the calculation of the Lagrange multiplier, which is widely interpreted as membrane tension (see detailed discussion in \cite{rangamani2014} and references therein), is an important aspect of the coupled physics.
By incorporating the viscous nature of the membrane, we ensure that the incompressibility constraint is met rigorously at all times and therefore obtain the tension fields on the membrane.
Our calculation of the heterogeneous tension fields are consistent with previous models as noted above and with experimental observations \cite{Shi2018-wa}. 
{Moreover, a lower tension inside the phase-separated domain further supports the existence of line tension at the domain boundary, which has been observed experimentally \cite{baumgart2003Nature}.} 

{Finally, we discuss the relevance of our model in the context of biological systems. 
Coarse-grained molecular dynamic simulations of N-BAR proteins on flat membranes and spherical vesicles showed that at low protein density these proteins form linear aggregates and meshes on the membrane surface \cite{simunovic2013linear}. 
Many proteins, especially those that belong to the coat family of proteins including clathrin and COP, are known to aggregate on the membrane and their aggregation results in morphologolical changes of the membrane \cite{sens2008biophysical}. 
The nucleation of these protein aggregates and the subsequent deformation of the membrane has been studied using simplified systems \cite{simunovic2016curvature}. 
While the exact role of lipid flow, diffusion, and aggregation is often not unraveled in these experiments, they have shown that the extent of curvature induced depends on multiple physical parameters including the composition of the membrane and the nature of the protein \cite{McMahon2005-fk,gallop2006mechanism}.
From a physiological perspective, many neurodegenerative diseases such as Alzheimer's disease, Parkinson's disease, and Huntington's disease are associated with surface aggregation of proteins in cells. 
Even though the precise mechanisms of such aggregation are not fully established, the role of membrane-protein interactions, particularly aggregation, is becoming increasingly important \cite{burke2013biophysical}. 

The formation of domains is not specific to lipid-protein systems but is also observed in vesicles that have two different kinds of lipids.
The temporal behavior of formation of disordered lipid domains was studied in a ternary mixture of fluid membrane \cite{saeki2006growth} and it was shown that in-plane flow was critical to the formation of such domains \cite{nitschke2012finite} and that smaller domains can be attracted towards larger domains following the internal flows \cite{yanagisawa2007flow}.
 }

In developing models for many of these experimental observations described above, aggregation of domains of protein-induced curvature is often assumed \textit{a priori} or curvature is proposed as an organizing factor to explain cellular observations and experiments in reconstituted systems \cite{Hassinger2017-wt,Liu2006-ma, Liu2009-sr, Alimohamadi2018-ca, Ma2021-vu, Rangamani2021-dv, Yuan2021-su}. 
By developing a general theoretical framework that accounts for the coupled effects of protein diffusion, aggregation, and curvature generation, we have eliminated the need for such strong assumptions and more importantly, demonstrated that the intricate interactions between these different physics can lead to different regimes of pattern formation and membrane deformations. 
These regimes can be tuned and controlled by different parameters, allowing for exquisite control of experimental design. 
In summary, the comprehensive model that we have developed here allows for a broader interpretation and understanding of membrane-protein interactions in a unifying framework.

\section{Acknowledgments}
This work was supported by NIH NIGMS R01-132106, ONR N00014-20-1-2469 to P.R. and NSF CBET-1705377 to D.S.

\clearpage

\appendix
\counterwithin{figure}{section}

{\LARGE\bfseries Electronic Supplementary Information}

\section{Model development}

\subsection{Stress tensor on a surface}
The stress tensor $\boldsymbol{\Sigma}$ represents the state of stress at any location of the membrane and includes both the in-plane normal and shear stresses as well as out-of-plane shear stress due to bending. 
Each column of the stress tensor $\boldsymbol{\Sigma}$ constitutes the traction vector on the curve drawn on the membrane, known as the stress vector, and is represented as
\begin{myequation}
\boldsymbol{\Sigma}^{\alpha}= N^{\alpha \beta} \boldsymbol{a}_{\beta}+ S^{\alpha}\boldsymbol{n},
\end{myequation}
where $\boldsymbol{N}$ is the surface stress tensor, $\boldsymbol{S}$ represents the shearing force due to bending, and $a_{\beta}$ (for $\beta=1,2$) represents the surface tangent vector normal to the curve. 
The local equilibrium of forces, in the tangential and normal directions, is given by \cite{Steigmann1999-su}
\begin{myequation}
\label{eqn:stress_bal}
\nabla \cdot \boldsymbol{N}-\boldsymbol{S} \cdot \boldsymbol{b}=0,
\end{myequation}
\begin{myequation}
\nabla \cdot \boldsymbol{S}+\boldsymbol{N}: \boldsymbol{b}+p=0, 
\label{eqn:stress_bal_normal}
\end{myequation}
with 
\begin{myequation}
\label{eqn:traction}
\boldsymbol{N}=\boldsymbol{\zeta}+\boldsymbol{\pi}+\boldsymbol{b} \cdot \boldsymbol{M} \quad \text { and } \quad \boldsymbol{S}=-\nabla  \cdot \boldsymbol{M	}.
\end{myequation}
Here, $\boldsymbol{\zeta}$ and $\boldsymbol{M}$ are the elastic stress and moment tensors, $\boldsymbol{b}$ is the curvature tensor, and $\boldsymbol{\pi}$ is the viscous stress tensor. 
The elastic stress and moment tensors can be obtained from the energy density for an incompressible membrane as \cite{Steigmann1999-su,mahapatra2020transport}
\begin{myequation}
\begin{split}
\label{eqn:stress}
\boldsymbol{\zeta} &=-2 \kappa(H-\ell \sigma) \boldsymbol{b}-2 \bar{\kappa} K \boldsymbol{a}-\xi \boldsymbol{a},  \\
\boldsymbol{M} &=\kappa(H-\ell \sigma) \boldsymbol{a}+\bar{\kappa}\left(2 H \boldsymbol{a}-\boldsymbol{b}\right), 
\end{split}
\end{myequation}
where $\xi$ is the Lagrange multiplier that imposes the incompressibility constraint and $\boldsymbol{a}$ 
is the metric tensor of the surface. 
The surface tension $\lambda$ is related to $\xi$ with the following expression \cite{Steigmann1999-su}
\begin{myequation}
\label{eqn:tension_lag}
\lambda= -(\xi+W).
\end{myequation}
The viscous stresses obey the constitutive relation \cite{rangamani2013interaction} 
\begin{myequation}
\label{eqn:vis_stress}
\boldsymbol{\pi}=2 \nu\left[\boldsymbol{d}-w \boldsymbol{b}\right].
\end{myequation}
Here, 
\begin{myequation}
\label{eqn:strain_rate}
\boldsymbol{d}=\left(\nabla  \boldsymbol{v}+\nabla  \boldsymbol{v}^{T}\right) / 2,
\end{myequation}
 is the rate-of-strain tensor expressed in terms of the 
velocity field $\boldsymbol{v}$
(see \cite{rangamani2013interaction,rangamani2014,mahapatra2020transport} for details). $w$ is the normal velocity of the surface, given by

\begin{myequation}
w= \boldsymbol{n}\cdot \boldsymbol{r}_t,
\end{myequation}
where $\boldsymbol{r}$ is the position vector of a material point on the surface. 

\subsection{Dimensionless governing equations}
Here we summarize the governing equations for the coupled dynamics of the system in the dimensionaless form.
The tangential force balance equation becomes
\begin{myequation}
\begin{split}
    \nabla \tilde{\lambda}-4 \tilde{w}  \nabla \tilde{H} &+ 2 ( \nabla  \cdot \tilde{\boldsymbol{d}}-\nabla \tilde{w} \cdot \tilde{\boldsymbol{b}}) \\
= &-\nabla  \phi \bigg[ \frac{2 \hat{B} \hat{S}}{\hat{T}} \log \frac{\phi}{1-\phi} - \frac{4 \hat{L}\hat{S}}{\hat{T}} (\tilde{H}-\hat{L} \hat{S} \phi) -\frac{\hat{A}\hat{B}\hat{S}}{\hat{T}} (2 \phi-1)- \frac{\hat{A}\hat{B}}{\hat{T}} \Delta  \phi\bigg], \label{eqn:tang_force_bal_nd}
\end{split}
\end{myequation}
along with the surface incompressibility relation,
\begin{myequation}
\label{eqn:cont_nd}
\nabla  \cdot \tilde{\boldsymbol{v}}=\,2 \tilde{w} \tilde{H}. 
\end{myequation}
The normal force balance relation takes the following form
\begin{myequation}
\label{eqn:norm_force_bal_nd}
\begin{split}
     \Delta & (\tilde{H}-\hat{L} \hat{S} \phi) +2(\tilde{H}-\ell L \sigma_s \phi)(2\tilde{H}^{2}-\tilde{K}) -2 \hat{B} \hat{S} \, \tilde{H}\bigg[ \{ \phi \log \phi  +(1-\phi) \log \left(1-\phi\right)\}+ \frac{\hat{A}}{2  } \phi (1 -\phi)   \\
     &  + \frac{\hat{A}}{4 \hat{S} }  |\nabla  \phi |^2 \bigg] -2\tilde{H} \left[(\tilde{H}-\ell \sigma_s L \phi)^2 + \frac{\bar{\kappa}}{\kappa} \tilde{K}\right]   - \hat{T} \big[ \tilde{\boldsymbol{b}}:\tilde{\boldsymbol{d}} -w(4\tilde{H}^{2}-2\tilde{K})\big] =\tilde{p}+\hat{T} \tilde{\lambda} \tilde{H}.
\end{split}
\end{myequation}
The mass conservation of proteins is given by
\begin{myequation}
\label{eqn:agg_diff_comp_nd}
\begin{split}
    \phi_t  + Pe & ~ \nabla  \cdot (\tilde{\boldsymbol{v}}\phi) =  \Delta \phi \bigg[  \frac{1}{1-\phi} + \frac{2 \hat{L} ^2 \hat{S} }{\hat{B}} \phi - \hat{A} \phi \bigg] -  \phi \bigg[ \frac{2 \hat{L}}{\hat{B}} \Delta H  + \frac{\hat{A}}{2 \hat{S}} \Delta^2 \phi\bigg]\\
    &+  \nabla  \phi \cdot  \bigg[  \nabla \phi \bigg(   \frac{1}{(1-\phi)^2} + \frac{2 \hat{L} ^2 \hat{S} }{\hat{B}} -\hat{A} \bigg)   - \frac{2 \hat{L}}{\hat{B}}  \nabla  \tilde{H} 
    -\frac{\hat{A}}{2 \hat{S}}  \nabla (\Delta \phi) \bigg].
\end{split}
\end{myequation}

\subsection{Governing equations in the linear Monge regime}

The continuity condition and tangential force balance simplify as
\begin{myequation}
    {{\nabla}}\cdot \boldsymbol{v}=2{w} {H},
       \label{eqn:LM_cont}
\end{myequation}
 and,
\begin{myequation}
\begin{split}
   \quad {\nabla} {\lambda} + &\nabla^2 {\boldsymbol{v}} + \nabla (\nabla \cdot {\boldsymbol{v}})- 4 {w} {\nabla} {H} -2 {\nabla}w : {\nabla} {\nabla} {z} =\\
&-{\nabla} \phi \bigg[  \frac{2 \hat{B} \hat{S}}{\hat{T}}   \log \frac{\phi}{1-\phi} -
 \frac{4 \hat{L} \hat{S}}{\hat{T}}(H-\hat{L}\hat{S} \phi)-\frac{\hat{A} \hat{B} \hat{S}}{\hat{T}} (2 \phi -1)- \frac{\hat{A} \hat{B}}{\hat{T}} \nabla^2 \phi \bigg].
   \label{eqn:LM_tang}
\end{split}
\end{myequation}
The normal force balance  \Cref{eqn:norm_force_bal_nd} reduces to
\begin{myequation}
\begin{split}
     &\nabla^4 {z}-2\hat{L}\hat{S}\nabla^2 \phi 
     -2  \hat{B} \hat{S}\nabla^2 {z}\bigg[\{\phi \log \phi +(1-\phi)\log (1-\phi)\} 
      +\frac{\hat{A}}{2} \phi (1-\phi)+ \frac{\hat{A}}{4 \hat{S}} |\nabla \phi|^2 
      +\frac{\hat{L}^2 \hat{S} }{\hat{B}} \phi^2\bigg]\\
    &-\hat{T} ({\nabla} \boldsymbol{v}+ {\nabla} \boldsymbol{v}^T): {\nabla}{\nabla} {z}={p}+\hat{T}{\lambda} \nabla^2 {z} .
       \label{eqn:LM_norm}
\end{split}
\end{myequation}
Finally, the transport equation for the protein density field \Cref{eqn:agg_diff_comp_nd} takes on the following form:  
\begin{myequation}
\begin{split}
    \phi_t  + Pe ~  {\nabla} \cdot ( \boldsymbol{v}\phi ) & =  \nabla^2 \phi \bigg[  \frac{1}{1-\phi} + \frac{2\hat{L}^2 \hat{S}}{\hat{B} } \phi - \hat{A} \phi \bigg]-  \phi \bigg[\frac{2\hat{L}}{ \hat{B}} \nabla^2 H  + \frac{\hat{A}}{2 \hat{S}} \nabla^4 \phi\bigg]\\ 
    & + \nabla \phi \cdot \bigg[ \nabla \phi \bigg(   \frac{1}{(1-\phi)^2} + \frac{2\hat{L}^2 \hat{S}}{\hat{B} } -\hat{A} \bigg)  
     - \frac{2\hat{L} }{ \hat{B}} \nabla H -\frac{\hat{A}}{2 \hat{S}} \nabla (\nabla^2 \phi) \bigg] .
       \label{eqn:LM_diff}
\end{split}
\end{myequation}

\subsection{Linear stability analysis in the linear Monge regime}
We substitute the follwoing normal modes into \Cref{eqn:LM_norm_lin} and \Cref{eqn:LM_diff_lin},   
\begin{myequation}
    \phi ^{\prime} = \Phi e^{\alpha t} e^{i 2 \pi \boldsymbol{k}\cdot \boldsymbol{x}} \quad \text{and} \quad z^{\prime} = Z e^{\alpha t} e^{i 2 \pi \boldsymbol{k}\cdot \boldsymbol{x}},
    \label{eqn:LM_normal_mode}
\end{myequation} 
yielding the relations
\begin{myequation}
\begin{split}
        & Z\bigg[ 16 \pi^4 k^4 + 8 \pi^2k^2 \hat{B} \hat{S}  \bigg( \{\phi_0 \log \phi_0 +(1-\phi_0)\log (1-\phi_0)\} 
       \\
      &+ \frac{\hat{A}}{2} \phi_0 (1-\phi_0) 
      +\frac{\hat{L}^2 \hat{S} }{\hat{B}} \phi_0^2\bigg) + 4 \pi^2 k^2 \hat{T} \bigg]  = -8 \pi^2 k^2 \hat{L} \hat{S} \Phi,
      \label{eqn:z_mode}
\end{split}
\end{myequation}
and
\begin{myequation}
\begin{split}
    \alpha  \Phi & =  -4\pi^2 k^2 \Phi\bigg[  \frac{1}{1-\phi_0} + \frac{2\hat{L}^2 \hat{S}}{\hat{B} } \phi_0 - \hat{A} \phi_0 \bigg]- 16 \pi^4 k^4 \phi_0 \bigg[\frac{\hat{L}}{ \hat{B}}  Z   + \frac{\hat{A}}{ \hat{2S}}  \Phi \bigg].
    \label{eqn:phi_mode}
\end{split}
\end{myequation}
Eliminating variables $Z$ and $\Phi$, we obtain the dispersion relation given in \Cref{eqn:LM_dispersion}.

\subsection{Numerical methods} 
We solved the dimensionless governing equations in the linear Monge regime (\Cref{eqn:LM_cont} to \Cref{eqn:LM_diff}) numerically inside a square domain with periodic boundary conditions.
Numerical simulations were performed on a spatial uniform grid of size 64 $\times$ 64 for the lower values of $\hat{S}$ (200 and 500). However, we considered a finer uniform grid of size 128 $\times$ 128 for the higher values of $\hat{S}$ (1000 and 2000), where we observed smaller sizes of protein aggregates.
We used a finite difference scheme to solve the transport equation for the protein density (\Cref{eqn:LM_diff}), whereas the velocity (\Cref{eqn:LM_cont} and \Cref{eqn:LM_tang}) and the shape (\Cref{eqn:LM_norm}) were solved using a Fourier spectral method \cite{hasimoto1959periodic,canutobook}. 
A semi-implicit scheme was used for the time marching for the protein density $\phi$ with a time step $\Delta t = 3 \times 10^{-4}$, where the nonlinear terms involving velocity and curvature were treated explicitly. In contrast, the nonlinear aggregation-diffusion terms were treated with linear implicit terms. The resulting transport equation is shown below
\begin{myequation}
\begin{split}
    &\frac{\phi^{n+1}-\phi^n}{\Delta t}  + Pe ~  {\nabla} \cdot ( \boldsymbol{v}^{\overline{n+1}}\phi^{n+1} )  =  \nabla^2 \phi^{n+1} \bigg[  \frac{1}{1-\phi} + \frac{2\hat{L}^2 \hat{S}}{\hat{B} } \phi - \hat{A} \phi \bigg]^{\overline{n+1}}-  \phi^{n+1} \bigg[\frac{2\hat{L}}{ \hat{B}} \nabla^2 H^{\overline{n+1}}\bigg]  \\ 
    & \quad \quad  + \phi^{\overline{n+1}} \bigg[\frac{\hat{A}}{2 \hat{S}} \nabla^4 \phi^{n+1}\bigg] + \nabla \phi^{n+1} \cdot \bigg[ \nabla \phi \bigg(   \frac{1}{(1-\phi)^2} + \frac{2\hat{L}^2 \hat{S}}{\hat{B} } -\hat{A} \bigg)  
     - \frac{2\hat{L} }{ \hat{B}} \nabla H -\frac{\hat{A}}{2 \hat{S}} \nabla (\nabla^2 \phi) \bigg]^{\overline{n+1}},
       \label{eqn:LM_diff_linz}
\end{split}
\end{myequation}
where the superscript $\overline{n+1}$ indicates the explicit terms for time step $n+1$, for which the currently available values were considered. The explicit terms were further updated using an iterative scheme, and within each iteration, velocity and shape were recalculated for the updated values of protein density.
The iterations were performed within a time step until convergence was achieved.
For the convergence within a time step, we used a tolerance of $5 \times 10^{-7}$. When the differences between values of variables from successive iterations fell below the tolerance, we considered the values of the variables to be converged in that time step.
The Fortran code for the numerical simulation is available on \href{https://github.com/armahapa/protein_aggregation_in_membranes}{\Blue{https://github.com/armahapa/protein\_aggregation\_in\_membranes}}.


\clearpage

\section{Supplementary figures}

\begin{figure}[htbp]
    \centering
    \includegraphics[width=\textwidth]{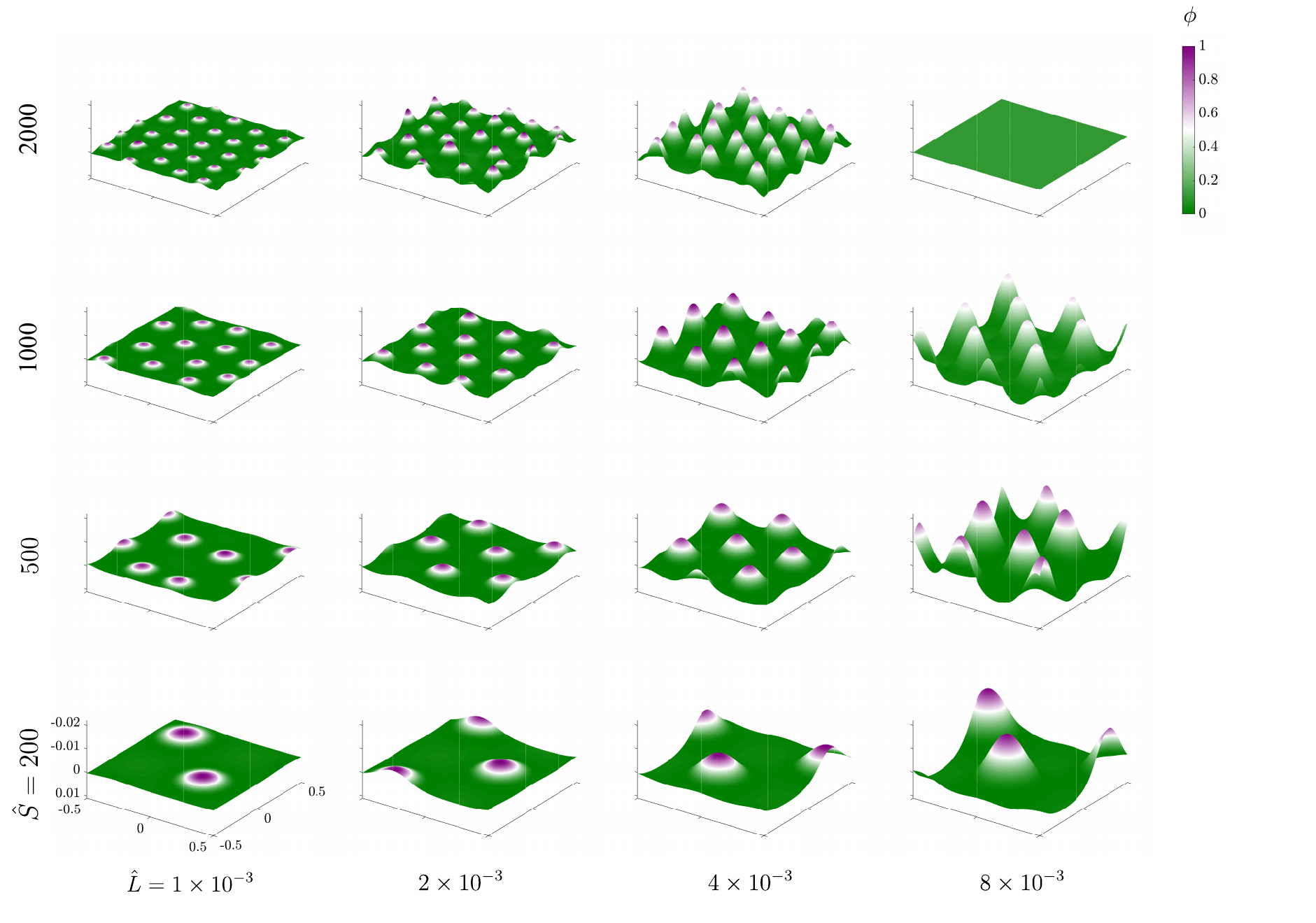}
    \caption{Protein distribution on the deformed membrane at a long time mimicking the steady state in the plane of $\hat{L}$ and $\hat{S}$, with $\hat{A}=25$.}
    \label{fig:LM_z_appendix}
\end{figure}

\begin{figure}[htbp]
    \centering
    \includegraphics[width=\textwidth]{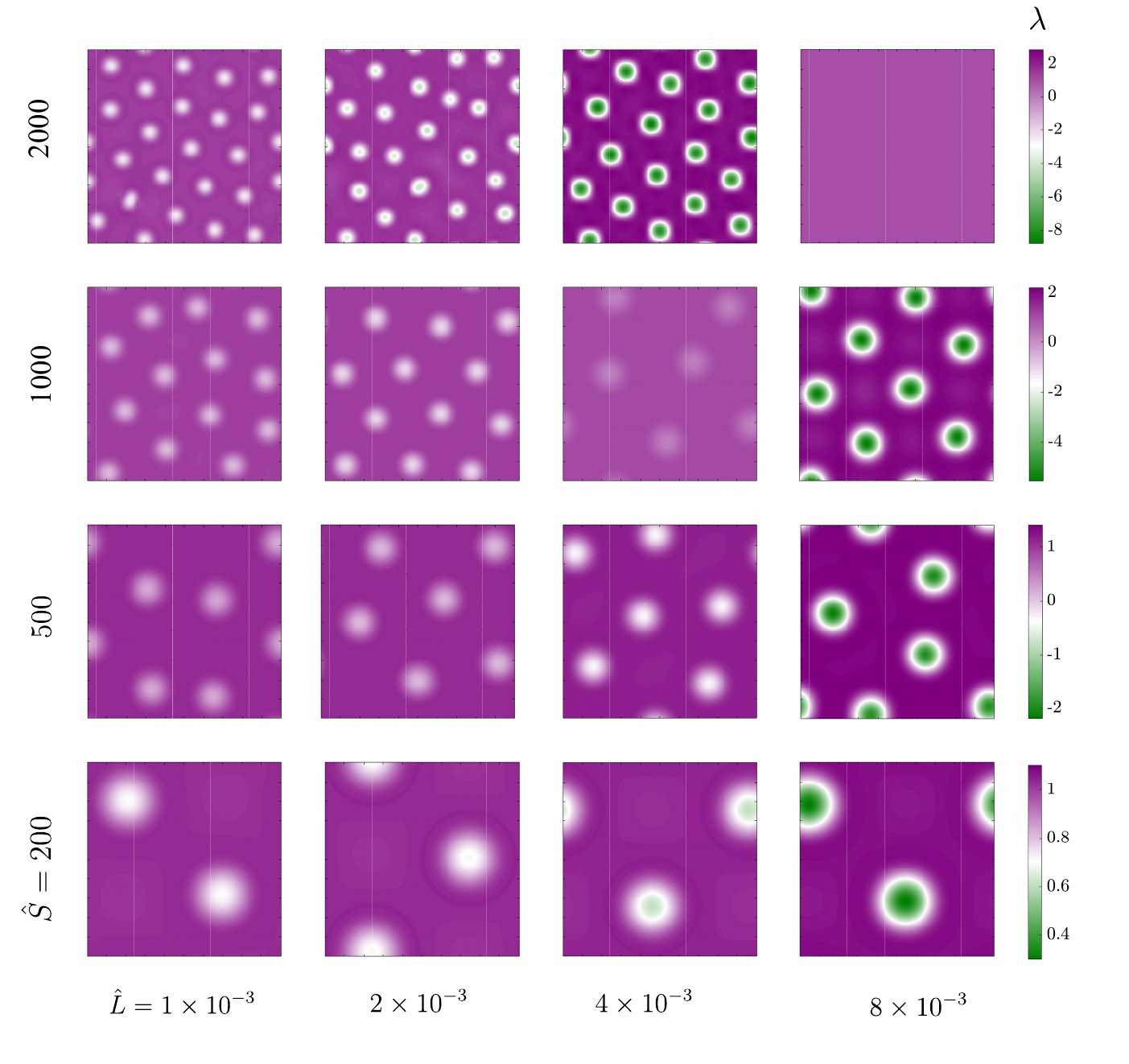}
    \caption{Membrane tension on the projected membrane surface at a long time mimicking the steady state in the plane of $\hat{L}$ and $\hat{S}$, with $\hat{A}=25$.}
    \label{fig:LM_Lam_appendix}
\end{figure}

\clearpage
\section{Supplementary movies}
\href{https://drive.google.com/file/d/1h-2WZgW5tpKKJssts3howy3HecwslYMH/view?usp=sharing}{\color{blue}{\textbf{Movie M1:}}} Time evolution of the protein density on a flat square membrane of area 1 $\mu$m for $\hat{A}=25$ and $\hat{S}=200$ in the Cahn-Hilliard case.

\href{https://drive.google.com/file/d/19jSHyU0_hfL6UMa6_ZddD-9Sk-I7IpvN/view?usp=sharing}{\color{blue}{\textbf{Movie M2:}}} Time evolution of the protein density on a flat square membrane of area 1 $\mu$m for $\hat{A}=25$ and $\hat{S}=500$ in the Cahn-Hilliard case.

\href{https://drive.google.com/file/d/10WcSacJUYPI2ThyplGgtQOk8QNOt6PVB/view?usp=sharing}{\color{blue}{\textbf{Movie M3}}} Time evolution of the protein density on a flat square membrane of area 1 $\mu$m for $\hat{A}=25$ and $\hat{S}=1000$ in the Cahn-Hilliard case.

\href{https://drive.google.com/file/d/1PLMvH0R5-PxBw7Ov7q5pnHyhfq5Zt46k/view?usp=sharing}{\color{blue}{\textbf{Movie M4:}}} Time evolution of the membrane deformation and protein density in a square membrane of size 1 $\mu$m$^2$ for $\hat{A}=25, \hat{S}=200$ and $\hat{L}=8 \times 10^{-4}$ for the fully coupled system.

\href{https://drive.google.com/file/d/1yb3EMkPSe5MNyMZVu6-DaTHH5Y_XJTvo/view?usp=sharing}{\color{blue}{\textbf{Movie M5:}}} Time evolution of the membrane deformation and protein density in a square membrane of size 1 $\mu$m$^2$ for $\hat{A}=25, \hat{S}=1000$ and $\hat{L}=8 \times 10^{-4}$ for the fully coupled system.

\href{https://drive.google.com/file/d/1-nQLsUQmjpLKy1f9Bod127AfelKqoVZz/view?usp=sharing}{\color{blue}{\textbf{Movie M6:}}} Time evolution of the membrane deformation and protein density in a square membrane of size 1 $\mu$m$^2$ for $\hat{A}=25, \hat{S}=2000$ and $\hat{L}=8 \times 10^{-4}$ for the fully coupled system.

 \clearpage
 \printbibliography


\end{document}